\documentclass[10pt,conference]{IEEEtran}
\usepackage{cite}
\usepackage{amsmath,amssymb,amsfonts}
\usepackage{algorithmic}
\usepackage{graphicx}
\usepackage{textcomp}
\usepackage{xcolor}
\usepackage[hyphens]{url}
\usepackage{fancyhdr}
\usepackage{hyperref}
\usepackage{algorithmic}
\usepackage[linesnumbered,lined,ruled,vlined]{algorithm2e}
\usepackage{bm}
\usepackage{pifont}
\usepackage{soul}
\usepackage{bbding}
\usepackage{wasysym}
\usepackage{titlesec}
\usepackage{pifont}
\usepackage[square, comma, sort&compress, numbers]{natbib}
\usepackage{threeparttable}
\usepackage{multirow}
 
\usepackage{marvosym}
\usepackage[table]{xcolor}
\usepackage{booktabs}
\usepackage{array}

\hypersetup{
  colorlinks=true,
  linkcolor=blue,
  urlcolor=blue,
  citecolor=blue
}

\definecolor{mygray}{gray}{.94}
\definecolor{Common}{rgb}{0.9765, 0.9373, 0.7490}
\definecolor{RA}{rgb}{0.8471, 0.8980, 0.9765}
\definecolor{RB}{rgb}{0.8569,0.8659, 0.8667}
\definecolor{RC}{rgb}{0.9333,0.6471,0.6510}
\definecolor{RD}{rgb}{0.0353,1.0000,0.6000}
\definecolor{RE}{rgb}{0.9882,0.65,0.45}
\definecolor{RF}{rgb}{0.9882,0.8471,0.8039}

\newcommand{\hpcayear}{2026}

\newcommand{\hpcasubmissionnumber}{94}
\title{WATOS: Efficient LLM Training Strategies and Architecture Co-exploration for Wafer-scale Chip}

\def\hpcacameraready{}

\newcommand\hpcaauthors{Huizheng Wang$^\dagger$, Zichuan Wang$^\dagger$, Hongbin Wang$^\dagger$, Jingxiang Hou$^\dagger$, Taiquan Wei$^\dagger$, Chao Li$^\ddagger$, \\ 
Yang Hu$^\dagger$\textsuperscript{\Letter}, Shouyi Yin$^\dagger$$^*$}
\newcommand\hpcaaffiliation{$^\dagger$School of Integrated Circuits, BNRist, Tsinghua University, Beijing, China, 100084 \\
$^\ddagger$School of Computer Science and Engineering, Shanghai Jiao Tong University, Shanghai, China, 200240 \\
$^*$Shanghai Artificial Intelligence Laboratory, Shanghai, China, 200433}
\newcommand\hpcaemail{\textsuperscript{\Letter}Corresponding author, hu\_yang@tsinghua.edu.cn}

\author{
  \ifdefined\hpcacameraready
    \IEEEauthorblockN{\hpcaauthors{}}
      \IEEEauthorblockA{
        \hpcaaffiliation{} \\
        \hpcaemail{}
      }
  \else
    \IEEEauthorblockN{\normalsize{HPCA \hpcayear{} Submission
      \textbf{\#\hpcasubmissionnumber{}}} \\
      \IEEEauthorblockA{
        Confidential Draft \\
        Do NOT Distribute!!
      }
    }
  \fi 
}

\fancypagestyle{camerareadyfirstpage}{%
  \fancyhead{}
  
  \fancyhead[C]{
    \ifdefined\aeopen
    \parbox[][12mm][t]{13.5cm}{\hpcayear{} IEEE International Symposium on High-Performance Computer Architecture (HPCA)}    
    \else
      \ifdefined\aereviewed
      \parbox[][12mm][t]{13.5cm}{\hpcayear{} IEEE International Symposium on High-Performance Computer Architecture (HPCA)}
      \else
      \ifdefined\aereproduced
      \parbox[][12mm][t]{13.5cm}{\hpcayear{} IEEE International Symposium on High-Performance Computer Architecture (HPCA)}
      \else
      \parbox[][0mm][t]{13.5cm}{\hpcayear{} IEEE International Symposium on High-Performance Computer Architecture (HPCA)}
    \fi 
    \fi 
    \fi 
    \ifdefined\aeopen 
      \includegraphics[width=12mm,height=12mm]{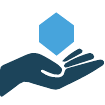}
    \fi 
    \ifdefined\aereviewed
      \includegraphics[width=12mm,height=12mm]{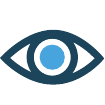}
    \fi 
    \ifdefined\aereproduced
      \includegraphics[width=12mm,height=12mm]{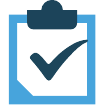}
    \fi
  }
  \fancyfoot[C]{}
}
\begin{document}
\maketitle

%Enables the camera ready header and footer
\ifdefined\hpcacameraready 
  \thispagestyle{camerareadyfirstpage}
  \pagestyle{empty}
\else
  \thispagestyle{plain}
  \pagestyle{plain}
\fi

\newcommand{\hpcaheight}{0mm}
\ifdefined\eaopen
\renewcommand{\hpcaheight}{12mm}
\fi

%\maketitle

\setcounter{page}{1}

\begin{abstract}
Training large language models (LLMs) imposes extreme demands on computation, memory capacity, and interconnect bandwidth, driven by their ever-increasing parameter scales and intensive data movement. Wafer-scale integration offers a promising solution by densely integrating multiple single-die chips with high-speed die-to-die (D2D) interconnects. However, the limited wafer area necessitates trade-offs among compute, memory, and communication resources. Fully harnessing the potential of wafer-scale integration while mitigating its architectural constraints is essential for maximizing LLM training performance. This imposes significant challenges for the co-optimization of architecture and training strategies. Unfortunately, existing approaches all fall short in addressing these challenges. 

To bridge the gap, we propose WATOS, a co-exploration framework for LLM training strategy and wafer-scale architecture. We first define a highly configurable hardware template designed to explore optimal architectural parameters for wafer-scale chips. Based on it, we capitalize on the high D2D bandwidth and fine-grained operation advantages inherent to wafer-scale chips to explore optimal parallelism and resource allocation strategies, effectively addressing the memory underutilization issues during LLM training. Compared to the state-of-the-art (SOTA) LLM training framework Megatron and Cerebras' weight streaming wafer training strategy, WATOS can achieve an average overall throughput improvement of $2.74\times$ and $1.53\times$ across various LLM models, respectively. In addition, we leverage WATOS to reveal intriguing insights about wafer-scale architecture design with the training of LLM workloads.

\end{abstract}

%\keywords{LLM training, wafer-scale chip, memory, recomputation, hardware}

\section{Introduction}
Large language models (LLMs) have emerged as pivotal components
in advancing AI in recent years \cite{dong2023towards,song2018situ}.  Their success is primarily attributed to unprecedented model scales, which impose substantial demands on computation, bandwidth, and memory resources \cite{wang2025beta,wang2025bitStopper}. Unfortunately, with the slowing of Moore’s Law \cite{moore1998cramming} and the limited photomask size \cite{wikichip_mask}, monolithic devices increasingly struggle to provide the necessary transistor density and memory capacity to support LLM training efficiently \cite{li2024large,han2024big}.

\begin{figure}[t]
\centering
\includegraphics[width=\linewidth]{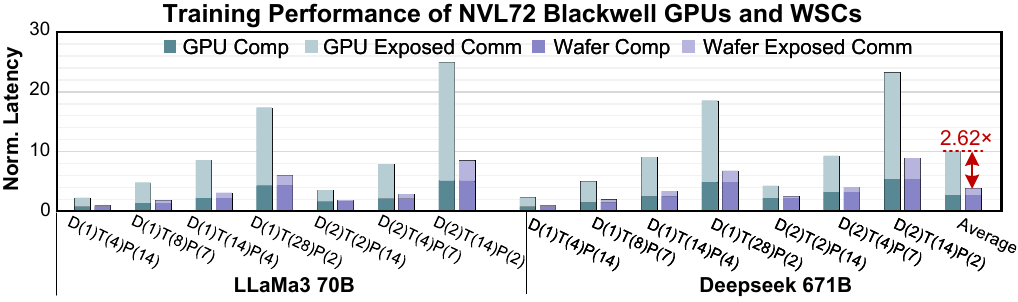}\vspace{-2mm}
\caption{Comparison of normalized training latency between NVL72 GB300 and WSCs under various parallelism configurations. The configuration D(1)T(4)P(14) indicates that there are 1/4/14 dies for DP/TP/PP dimension.}
\label{fig:Exposed Analysis}
\end{figure}

Wafer-scale chips (WSCs) \cite{hu2024wafer}, using advanced packaging technologies such as Chip-on-Wafer-on-Substrate (CoWoS) \cite{huang2021wafer} to scale up the chip area, delivering a promising solution to overcome these limitations and enable continuous transistor integration. Currently, UCLA's wafer-scale processor \cite{pal2021designing}, wafer-scale GPU \cite{pal2019architecting}, Tesla Dojo \cite{talpes2022dojo},  Cerebras WSE series \cite{lie2022cerebras,ltaief2023scaling,lie2024inside} and TSMC SoW-X \cite{shih2025sow} have been introduced. Compared to the most advanced GPUs \cite{nvidia_h1} with high-speed rack-scale fabrics (i.e., NVLinks \cite{nvidia_gb2}), these WSCs can provide around $6\times$ the inter-chip bandwidth and $5\times$ lower link latency. Our characterization in Fig. \ref{fig:Exposed Analysis} reveals that, under equivalent compute power, the 56-die WSC achieves a $2.62\times$ reduction in effective communication latency compared with the state-of-the-art (SOTA) 56-GPU NVIDIA GB300 system \cite{nvidia_gb3} configured with NVL72. This result highlights the substantial potential of WSCs for  LLM training.

However, unlike traditional ASIC designs \cite{moon2025t,wang2021efficient,kim2025slim,dong202528nm,wang2023efficient}, chiplet-based architectures \cite{park2025leveraging,cai2024gemini,shao2019simba,wang2022application} or DGX-class systems \cite{nvidia_dgx_gh200_whitepaper,tirumala2024nvidia,vspetko2021dgx}, WSCs face a unique architectural challenge: the total physical area available for compute and memory resources is strictly limited by the wafer's size, typically around 40,000\,mm$^2$. This constraint imposes a fundamental trade-off: Expanding on-wafer memory capacity necessarily reduces the silicon budget available for compute resources, and vice versa.

% Among these, a widely used, and high-yield wafer-scale approach is integrating neural processing units (NPUs) and DRAM chiplets together. 

% The significant increase in bandwidth dramatically reduces communication overhead during LLM training, particularly in large tensor parallel group, as illustrated in Fig. \ref{fig:Exposed Analysis}, thereby effectively improving throughput and giving WSCs a distinct advantage over traditional training systems.

% Typically, wafer-scale systems utilize high-density 3D integration \cite{hu2024wafer}. Vertical integration of modules such as power supply, cooling, and I/O interfaces achieves exceptional integration density, which in turn maximizes compute density and interconnect bandwidth. 

%带宽的大幅提升为LLM训练降低了大量通信时间，如Fig. \ref{}, 尤其是在，从而能够有效地提高吞吐率，使得wafer-scale chips具有传统的训练系统不具备的优势。

% Among these, one widely used, cost-effective, and high-yield wafer-scale approach is integrating NPU and DRAM chiplets together.

\textbf{Therefore, enabling efficient LLM training on WSCs demands tightly coordinated optimization along two fundamental levels}: \textbf{(1)} \textbf{\textit{Architecture level}}, where the system must be tailored to LLM workload characteristics, balancing compute and memory resources for optimal performance. \textbf{(2) }\textbf{\textit{Training strategy level}}, which involves designing optimal parallelism to align with architectural constraints, efficiently distributing tasks to fully exploit the WSC’s capabilities.

However, upon re-examining existing works, we identify that the co-design of training strategies and architectural levels for LLM training remains largely unexplored. Most existing research continues to focus on single-level optimizations. 

%\textbf{Single-level optimizations  on training strategy level}:

\textbf{Isolated optimizations on training strategy level}. Prior arts \cite{megatron_lm,shoeybi2019megatron,narayanan2021efficient,korthikanti2023reducing,sun2024adapipe,huang2025netzip} have proposed various training optimizations, such as auto parallelism, mixed precision \cite{micikevicius2017mixed,he2022campo}, offloading \cite{zhou2023mpress}, compression \cite{huang2025netzip}, and recomputation \cite{sun2024adapipe}, which significantly improve GPU training performance, as shown in the step 1 of Fig. \ref{fig:Contribution}. However, the performance ceiling is fundamentally constrained by the limited bandwidth of the board-level interconnect in DGX systems. Achieving higher performance urgently requires training strategies specifically designed for WSCs. 

% Step 1: Megatron可以在XX配置下使用XX时间训练XX模型
% Step 2: 然而，将Megatron直接应用到相同算力晶圆级芯片上需要XX倍的时间
% Step 3&4: 简单应用XX(粗暴地应用DSE中的方法)中的方法只有XX的提升
% Step 5: 应用了WATOS后相对GPU-Megatron和Wafer-Megatron有XX倍的提升

\textbf{Isolated optimizations on architecture level}. Several studies \cite{pal2019architecting,pal2021designing,chenwaferscale,feng2024switch,rashidi2024fred,yang2025pd} have attempted to explore wafer-level architectures, involving interconnect topologies \cite{yang2025pd}, interconnect substrates \cite{pal2019architecting}, and chiplet design \cite{chenwaferscale}. However, these works, on one hand, lack the collaborative optimizations with training strategies, leading to suboptimal performance, as step 2 in Fig. \ref{fig:Contribution}. For example, when applying Megatron's parallel setup to train Llama2-30B on a WSC, we observe a $80\%$ performance gap between real and potential performance. Therefore, to fully unlock the potential of hardware resources, it is critical to co-design training strategies with the architecture. As revealed in step 3\&4 of Fig. \ref{fig:Contribution}, the absence of training strategy-architecture synergy results in the inability to fully leverage the WSC architecture's potential. On the other hand, existing WSC designs often fail to thoroughly consider physical area constraints, thereby preventing the realization of an optimal wafer architecture design. 

% Slight variations in memory can lead to up to 20\% performance degradation, and change in the total number of dies may render the parallelization strategy completely inapplicable.

However, it is far from trivial to develop a co-design framework that integrates training strategy and architecture. It faces three key challenges: 1) The lack of configurable wafer-scale hardware templates for design space exploration (DSE). 2) The absence of systematic memory-centric training optimizations tailored to WSCs. 3) Naïve greedy co-design strategies are prone to getting trapped in local optima.  

To address these challenges, in this paper, we develop a physical-constraints aware training strategy and architecture co-exploration framework, WATOS, as the step 5 in Fig. \mbox{\ref{fig:Contribution}}. Our contributions are summarized as follows:

\begin{figure}[t]
\centering
\includegraphics[width=\linewidth]{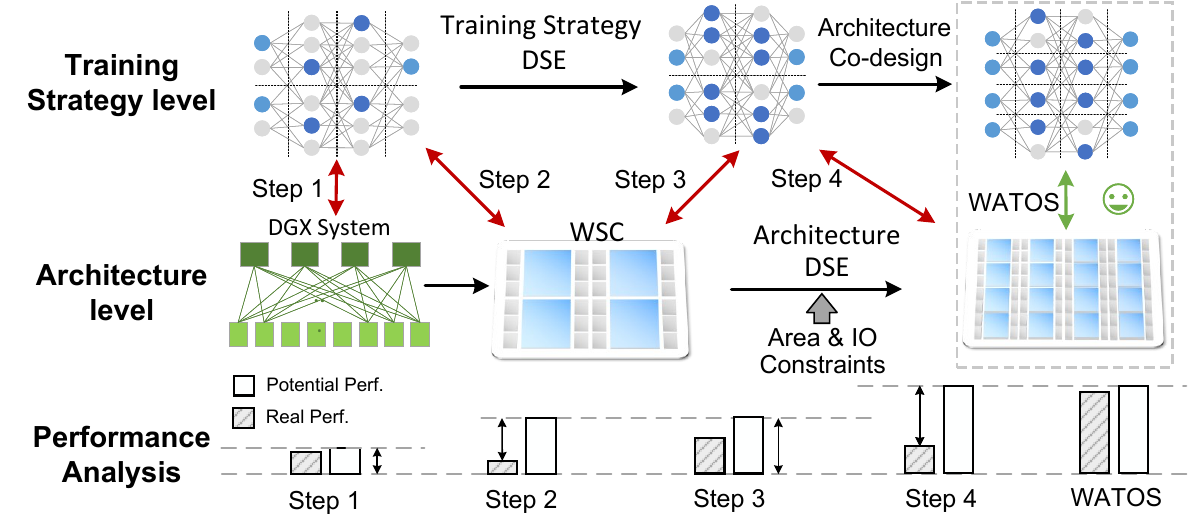}\vspace{-2mm}
\caption{Illustration of the training strategies and architecture co-design. The GPU system consists of 8 Blackwell Ultra GPUs, providing a total compute capacity of 40,000 TFLOPS, 2304 GB of HBM, 1.8 TB/s of NVLink interconnect bandwidth, and 8 TB/s of memory bandwidth. For fairness, the WSC is configured with equivalent compute power, as detailed in \S\ref{subsec:Overall_Perf}.}
\label{fig:Contribution}
\end{figure}

% The above analysis highlights that leveraging WSCs for efficient LLM training introduces intricate trade-offs and system-level challenges. To address these challenges, we make the following contributions: 

\textbf{1)} Targeting WSCs, we design a highly configurable hardware template. Leveraging this template, WATOS systematically identifies and exploits latent co-optimization opportunities between \textbf{wafer architecture} and \textbf{training strategies}. These opportunities become particularly critical in the post-Moore era. To the best of our knowledge, this is \textbf{the first work} to enable such joint optimization.

%体现先协调整个wafer的memory，在协调stage内部的memory
\textbf{2)} To efficiently explore the co-design space, WATOS integrates the following key innovations: \textbf{(1)} An early-pruning central scheduler to accelerate the DSE process. \textbf{(2)} A globally coordinated memory-efficient recomputation (GCMR) strategy that maximizes memory savings while minimizing pipeline bubble overhead. \textbf{(3)} A global communication-cost aware resource placement and allocation strategy to reduce communication costs. \textbf{(4)} A genetic algorithm (GA)-based optimizer to accelerate convergence toward the global optimum. 

% We specifically develop strategies tailored to the characteristics of WSCs to explore optimal model parallelism, resource allocation, placement, and memory management for LLM training. These algorithms works jointly to explore the most effective mapping and memory scheduling strategies, maximizing the utilization of hardware resources on WSCs.

% aim to explore the most effective mapping and memory scheduling strategies while efficiently managing the model data placement, 

\textbf{3)} Using WATOS, we uncover several intriguing insights for the WSC architecture and the LLM training: (1) With WATOS' joint optimization, WSCs with moderate per-die DRAM capacity can achieve a balanced allocation of compute, communication, and memory resources, yielding near-optimal training throughput. (2) With tailored resource allocation and memory scheduling, smaller TP configurations on WSCs can realize better training performance than larger ones. (3) Enabled by our dedicated memory management strategies, efficient LLM training on WSCs demands not only high DRAM bandwidth but also ample communication bandwidth, making their balance crucial for optimal performance. 

%(4) Our experiments on the impact of different optimization strategies reveal that the memory optimization techniques in WATOS contribute more significantly to large-scale models. 

\textbf{4)} Compared to the SOTA LLM training systems and Cerebras wafer training strategy, WATOS's co-optimized architecture and training strategies achieve an average performance gain of $2.74\times$ and $1.53\times$, respectively. 

\begin{figure}[t]
\centering
\includegraphics[width=\linewidth]{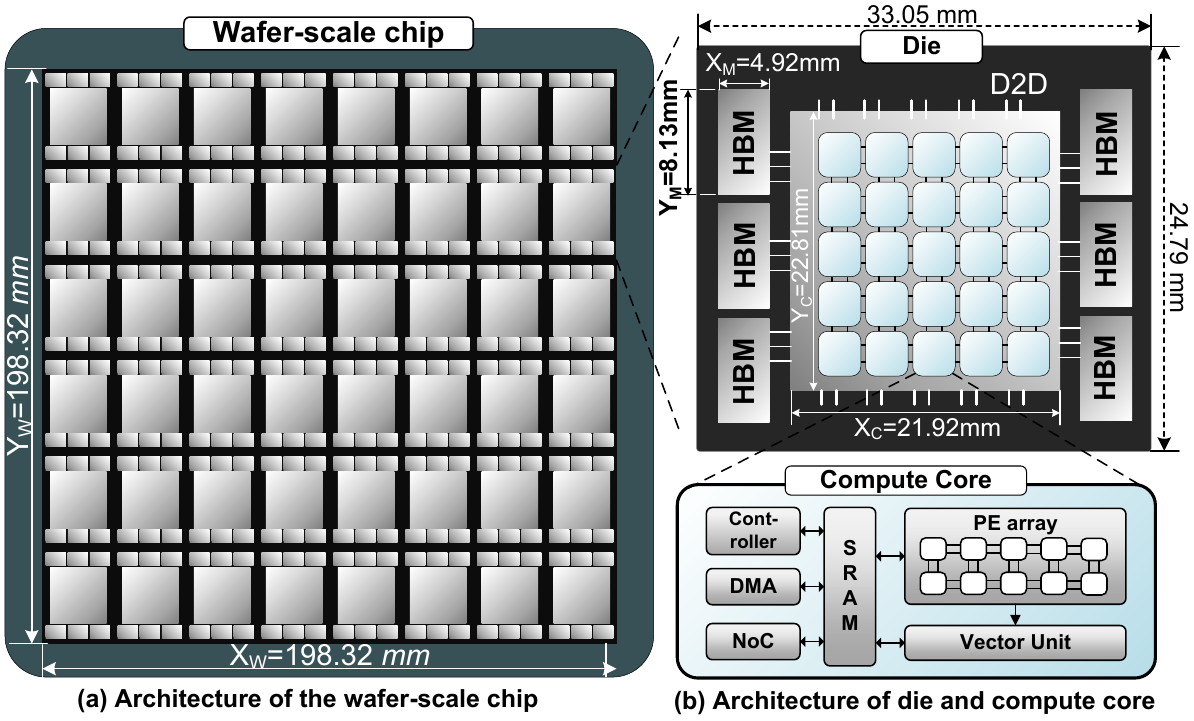}\vspace{-2mm}
\caption{Architecture of wafer-scale chips (WSCs).}
\label{fig:wafer}
\end{figure}

\section{Background}\label{sec:background}
\subsection{Wafer-scale Chip (WSC)}\label{subsec:wafer-scale-chip}
WSCs integrate multiple functional dies using advanced packaging technologies like TSV \cite{jiao2024low, guo2022review, lau2023recent}, CoWoS \cite{ huang2021wafer, hu2023cowos, yang2024signal}, and RDL \cite{lau2021hybrid, hou2023supercarrier, lau2020panel, hsia2024wafer}. There are two main approaches to wafer-scale integration: 1) Monolithic integration, where all compute dies and interconnects are fabricated directly on a single wafer substrate \cite{Cerebras2022Cerebras,lie2022cerebras,he2025waferllm}. 2) Chiplet-based integration, where compute dies are fabricated separately and bonded to the wafer-scale interconnect \cite{shao2019simba,pal2021designing,pal2019architecting,bajwa2018demonstration}. Given the latter offers greater flexibility, scalability, and yield, as individual dies can be independently optimized and tested before integration, this paper focuses on the chiplet-based WSC architecture. 

% \textbf{(Overall wafer architecture.} As illustrated in Fig.\ref{fig:wafer}, the hardware template adopts a three-level hierarchy: wafer, die and core. A 2D mesh topology is employed on the $198$mm $\times$ $198$mm, where adjacent dies communicate via die-to-die (D2D) interconnects. This topology supports multiple communication types, including Die-to-Die, Die-to-DRAM, DRAM-to-Die. While alternative topologies such as torus or fully-connected designs are possible, practical constraints on interconnect length and signal integrity often degrade bandwidth and efficiency. Therefore, the mesh topology remains the most effective architecture choice on wafer.
\textbf{Wafer-level architecture.} As illustrated in Fig. \ref{fig:wafer}, the WSC hardware template is structured into three hierarchies: wafer, die, and core. At the wafer level, a 2D mesh topology is deployed on the 198mm$\times$198mm wafer to achieve efficient communication between adjacent dies, supporting modes including Die-to-Die and Die-to-HBM. This topology offers superior bandwidth and signal integrity compared to alternatives, making it the preferred choice for wafer-scale architectures.

% \textbf{Die level.} Each die integrates both computation and DRAM memory, which marks a significant departure from conventional chiplet designs \cite{shao2019simba,cai2024gemini,tan2021nn,naffziger20202}, which typically decouple these functionalities. As depicted in Fig. \ref{fig:wafer}, each die consists of multiple computing cores, on-die NoC interconnections, memory (HBM chiplets) and D2D interconnect interfaces. The on-die NoC also adopts a 2D mesh topology, which enables high-speed core-to-core communication. The placement of D2D interfaces around the die periphery enables a scalable architecture.
\textbf{Die \& Core architecture.} Each die integrates both computation and DRAM resources, unlike conventional chiplet designs that typically decouple these components \cite{shao2019simba,cai2024gemini,tan2021nn,naffziger20202}.  As depicted in Fig. \ref{fig:wafer}, each die consists of multiple compute cores, a Network-on-Chip (NoC), HBM chiplets and peripheral D2D interfaces. The NoC adopts a 2D mesh topology, enabling low-latency communication across cores. At the core level, each core is managed by a local controller and communicates via the NoC and DMA engine. It comprises a shared SRAM, a PE array for GEMM operations, and a vector unit for scalar computation.

\textbf{Configurable Hardware Template.} Our hardware template offers a range of adjustable architectural parameters. These parameters include: 1) at the wafer level, compute die size ($X_C, Y_C$), DRAM die size ($X_M,Y_M$), the number of dies in horizontal ($N_D^X$, e.g., 8 in Fig. \ref{fig:wafer}), vertical dimensions ($N_D^Y$, e.g., 6 in Fig. \ref{fig:wafer}) along with DRAM/SRAM capacity, as well as NoC topologies. At the die level, compute units are configurable in core array dimensions and the number of MAC units per PE array. The PE microarchitecture and dataflow are derived from prior accelerator designs \cite{chen2016eyeriss,wang2024sofa,hegde2021mind,wang2025lapa,kao2020confuciux,wang2025mcbp,kwon2019understanding,tan2021nn,wang2018low,chen2014diannao,yang2020interstellar,wang2021efficientMIMO}. 

\begin{figure}[t]
\centering
\includegraphics[width=\linewidth]{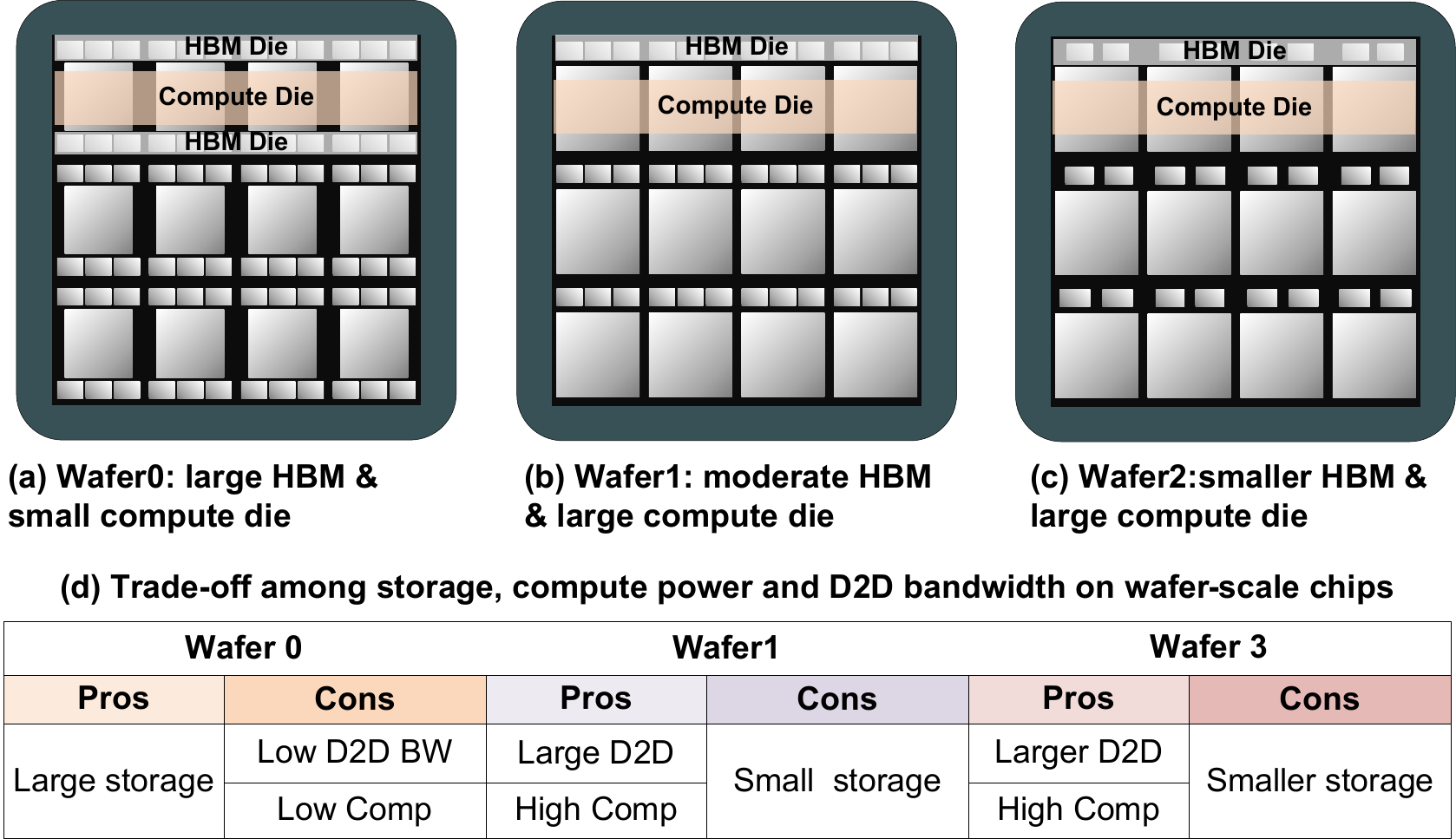}\vspace{-2mm}
\caption{\!\!Trade-off among storage, compute power and D2D bandwidth on WSC.}
\label{fig:wafer-trade-off}
\end{figure}

\subsection{Large Model Distributed Training}\label{subsec:Large_model_training}
There are three primary parallel training methods: data parallelism (DP) \cite{rajbhandari2020zero,xing2015petuum,ren2021zero}, where each die stores the full model and synchronizes gradients across dataset shards; tensor parallelism (TP) \cite{shoeybi2019megatron,wang2019supporting,dean2012large}, which partitions tensors along specific dimensions across dies; and pipeline parallelism (PP) \cite{huang2019gpipe,narayanan2021memory,fan2021dapple}, which splits the model into sequential stages executed in order. PP is often implemented using one-forward-one-backward (1F1B) schedule \cite{fan2021dapple}, which effectively reduces memory consumption compared to traditional PP strategies \mbox{\cite{huang2019gpipe,petrowski1993performance}}. As in Fig. \ref{fig:pipeline} (a), the 1F1B schedule consists of three phases: warmup, steady and ending. For $p$ stages and $n$ micro-batches ($p$$=$3, $n$$=$5 in Fig. \ref{fig:pipeline} (a)), a die at stage $s$ first performs $p$$-$$s$ forward passes during warmup, then alternates $n$$-$$p$$+$$s$ forward and backward passes in the steady phase, and completes $p$$-$$s$ backward passes in the ending phase. However, 1F1B has imbalanced memory usage \cite{sun2024adapipe}, as dies at the stage $s$ must retain intermediate activations for $p$$-$$s$ micro-batches, as shown by the number of saved activation blocks in Fig. \ref{fig:pipeline} (a).

%The scheduling process of 1F1B can be divided into three phases: warmup, steady, and ending phases. Given $p$ stages and $n$ micro-batches ($p=3$, $n=6$ in Fig. \ref{fig:1F1B}), dies in stage $s$ will first perform the forward passes of $p-s$ micro-batches in the warmup phase, then process $n-p+s$ backward and forward passes in the steady phase, and finally compute the backward passes of $p-s$ micro-batches in the ending phase. However, the 1F1B mechanism features ill-balanced memory occupancy, since dies working in the stage $s$ need to save the intermediate results of $p-s$ micro-batches, as depicted by the number of saved activation blocks in Fig. \ref{fig:1F1B}. 

Recently, bidirectional PP schedules \cite{deepseekai2025deepseekv3technicalreport,li2021chimera,liu2023hanayo} are proposed. However, these methods intensify memory pressure as model parameters are duplicated across stages. In addition, when $n$$>$$p$, these schedules often incur more pipeline bubbles than 1F1B. Moreover, DP also suffers from high memory overhead due to full model replication. Even advanced DP method like FSDP \cite{zhao2023FSDP}, which shards model states to reduce memory usage, perform poorly on WSCs due to heavy traffic of weights, gradients and optimizer states. As Fig. \mbox{\ref{fig:FSDP_Offloading}} (a) shows, such traffic congests the 2D-mesh NoC, causing a $20\%$–$40\%$ drop in bandwidth utilization compared to TP, which transmits activations only. \textbf{Thus, we focus on coordinating TP and PP under 1F1B scheduling to scale LLM training on WSCs.}

 \begin{figure}[t]
\centering
\includegraphics[width=\linewidth]{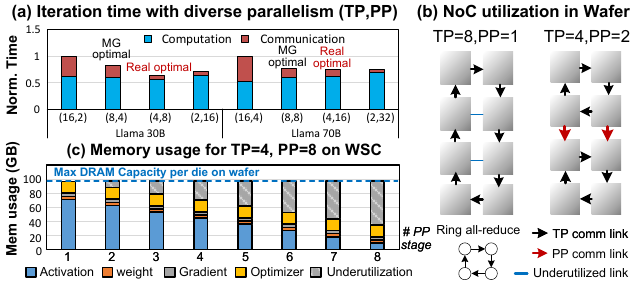}\vspace{-3mm}
\caption{Current LLM training parallelism framework is unsuitable for WSCs.}
\label{fig:TP_PP_Motivation}
\end{figure}

\begin{figure}[t]
\centering
\includegraphics[width=\linewidth]{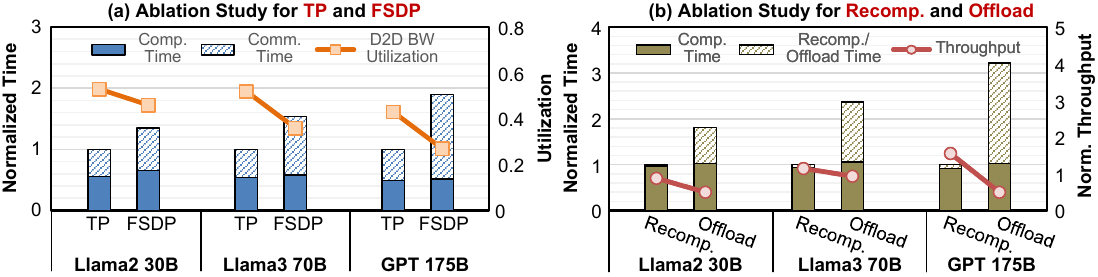}\vspace{-3mm}
\caption{Ablation Study for (a) FSDP and (b) Offloading. WSC is configured as \S\ref{subsec:set_up} with host-WSC PCIE 160GB/s \cite{talpes2022dojo}.}
\label{fig:FSDP_Offloading}
\end{figure}

\subsection{LLM Training System}

Batching is widely adopted in modern LLM training to improve computation efficiency and accelerate convergence \cite{nguyen2024memory,touvron2023llama,touvron2023llama2}. However, large batch sizes lead to excessive memory occupancy. To address this, training frameworks like Megatron-LM \cite{shoeybi2019megatron} and DeepSpeed \cite{rasley2020deepspeed} adopt memory optimization techniques such as mixed precision \cite{micikevicius2017mixed,kalamkar2019study,burgess2019bfloat16}, offloading \cite{ren2021zero,rajbhandari2021zero,pudipeddi2020training}, and activation recomputation \cite{chen2016training,patil2022poet,jain2020checkmate}. While effective in conventional systems, some of these are ill-suited for WSCs. For example, compared to the immense computation capability and on-chip bandwidth of WSCs, limited bandwidth between host and WSC leads to severe computation stalls, making offloading impractical in wafer scenarios. In contrast, \textbf{activation recomputation is more well-suited to WSCs, as it reduces memory usage without external data transfer}. As shown in Fig. \mbox{\ref{fig:FSDP_Offloading}} (b), compared with recomputation, offloading incurs substantially higher overhead, resulting in an average 2.2$\times$ increase in wall-time latency. Despite the advantage, the recomputation alters the memory-to-compute ratio, reshaping hardware resource demands. This calls for a co-design of training strategies and wafer architecture, an aspect that remains unexplored in existing works.

% To summarize, while current optimizations can alleviate the memory usage during training, systematic challenges such as imbalanced memory usage, inefficient resource allocation, and underutilized hardware persist, making them to fall short in wafer-scale chips. 

%\textbf{Recomputation}. To alleviate the memory occupancy during training, recomputation is proposed \cite{chen2016training}, which discards activations generated during forward propagation and recomputes them in the backward pass. However, while recomputation can effectively reduce the memory usage, it comes with the expense of increased computation burden. In this paper, by convention, we refer to an activation as a \emph{checkpoint} if it is not subject to recomputation.%

% To alleviate the memory usage during training, recomputation is proposed, which discards activations generated during forward propagation and recomputes them in the backward pass. However, while recomputation can effectively reduce the memory usage, it comes with the expense of increased computation burden. In this paper, by convention, we refer to an activation as a \emph{checkpoint} if it is not subject to recomputation.

\begin{figure}[t]
\centering
\includegraphics[width=\linewidth]{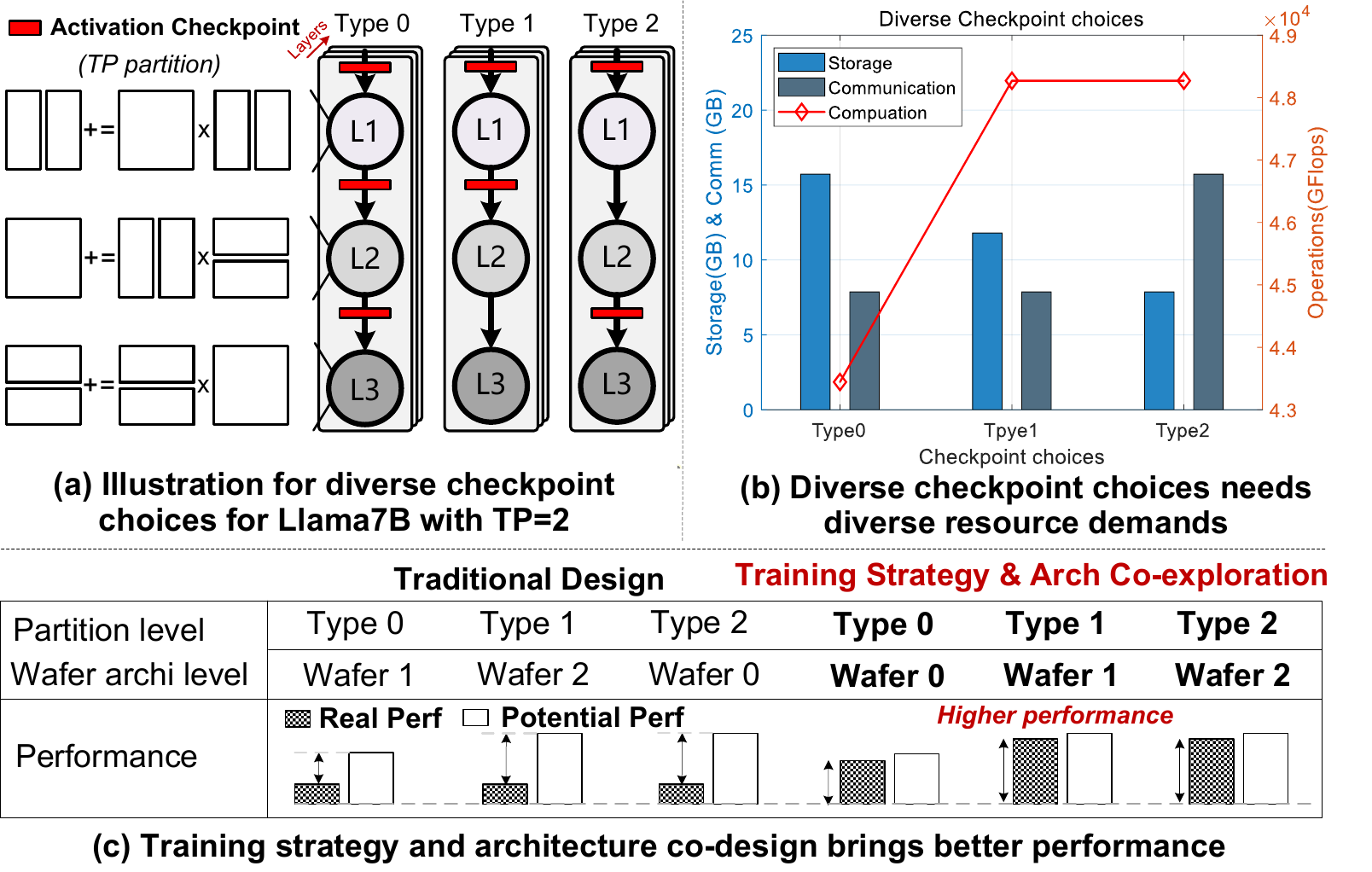}\vspace{-2mm}
\caption{Potential of training strategy and architecture co-design in WSCs.}
\label{fig:co-design}
\end{figure}

\section{Motivation}\label{sec:Motivation}
\subsection{Problems in LLM Training on WSCs}\label{subsec:Problem_in_LLM}
While existing training systems have recognized the resource-intensive characteristic of LLM training and adopted hybrid parallelism to address it, several key problems remain:

\textbf{First, existing systems often fail to achieve optimal parallelism on WSCs due to a lack of consideration for the underlying interconnect topology.} To demonstrate this, we conducted experiments using Llama-30B/70B models across 32 and 64 dies on a WSC. The compute die configuration is detailed in \S\ref{subsec:Archi_DSE}. Based on the model configurations, we derived the optimal parallelism settings recommended by the Megatron framework \cite{megatron_lm} (MG-optimal), which suggests (TP, PP) = (8, 4) for 32 dies and (8, 8) for 64 dies. We then deployed these settings on the WSC to evaluate actual performance. As shown in Fig. \ref{fig:TP_PP_Motivation} (a), on the WSC, configurations of (4, 8) for 32 dies and (4, 16) for 64 dies outperform the MG-optimal settings. This discrepancy arises from Megatron’s ignorance of the WSC's 2D mesh interconnect. As depicted in Fig. \ref{fig:TP_PP_Motivation} (b), naively setting TP = 8 leads to link under-utilization during Ring All-Reduce, affecting communication efficiency. In contrast, TP=4 with PP=2 achieves more balanced link utilization and better overall performance.

\textbf{Second, imbalanced memory usage incurred by large PP leads to underutilized DRAM resources on WSCs.} Fig. \ref{fig:TP_PP_Motivation} (c) shows peak DRAM usage during Llama-30B training with TP=4 and PP=8 on a WSC, where each die is equipped with 96 GB of DRAM (detailed in \S\ref{subsec:Archi_DSE}). The results reveal pronounced disparities in average DRAM utilization among dies on the WSC. Dies at the early stages of the pipeline exhibit much higher memory pressure than those at the tail. This skew is caused by checkpointed intermediate activations, which account for over $70\%$ of total memory usage. %Consequently, significant recomputation is incurred to alleviate memory pressure at early stages when training large models on WSCs, yet it will lead to extra overhead and severe pipeline bubbles \cite{fan2021dapple}, as illustrated in Fig. \ref{fig:pipeline} (a). Moreover, the naive recomputation strategy focuses on individual devices while lacking a global perspective, which results in additional computational overhead and underutilized memory, rendering inefficient training process, as depicted in Fig. \ref{fig:pipeline} (c). Hence, balancing memory utilization is essential for reducing recomputation cost and eventually reducing training iteration time.

\subsection{Trade-off among Compute-Memory-Comm on WSCs}\label{subsec:Trade-off in Compute-Memory-Comm}
%LLM trainings impose significant demands on computation, memory, and communication resources. While wafer-scale chips (WSC) effectively scale monolithic devices, they remain constrained by the limits of the wafer size. For example, the maximum usable area of a 12-inch wafer is approximately 39204 mm$^2$ (198mm $\times$198 mm). Therefore, WSCs necessitate a trade-off among the main components on wafer. However, existing research has not addressed this challenge. In this section, we discuss the trade-off among compute, memory and communication resources on WSCs.

LLM training imposes substantial demands on compute, memory, and communication resources. WSCs enable large-scale integration for diverse function dies, but remain constrained by the physical wafer size, which provide around 40,000 mm$^2$ of usable area on a 12-inch wafer. \textbf{This architecture feature highlights a trade-off among on-wafer compute, memory, and interconnect resources}.

% In this section, we discuss the fundamental trade-off among compute, memory and communication resources on WSCs.

%Fig. \ref{fig:wafer-trade-off} (a)-(c) depicts three types of DRAM chiplets configurations with a single compute die, corresponding to different DRAM capacities, respectively. We denote them as wafer 0, 1, 2, respectively. In wafer0, six DRAM chiplets are positioned on both sides of the compute chip, while in wafer1, they are arranged on a single side, with the number reduced to three. The saved DRAM space is then utilized to increase the area of the compute die. For wafer2, the compute die remains identical to those in wafer1, but the number of DRAM chiplets is reduced. This leaves more peripheral IO interfaces available for D2D interconnects. In other words, compared to wafer1, wafer2 provides higher D2D bandwidth.

\begin{figure}[t]
\centering
\includegraphics[width=\linewidth]{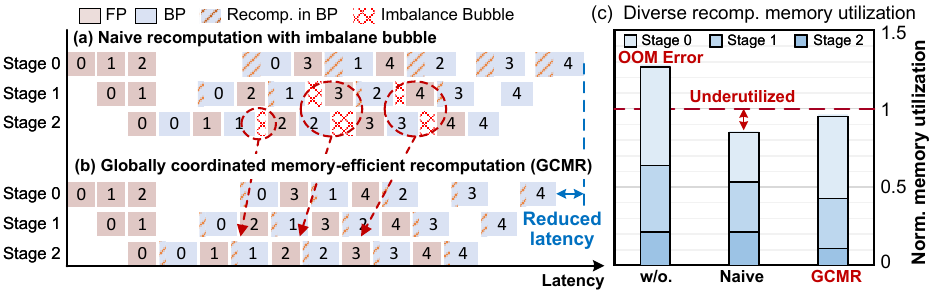}\vspace{-4mm}
\caption{Illustration of the imbalance bubble caused by naive recomputation and its underutilized memory.}
\label{fig:pipeline}\vspace{-2mm}
\end{figure}

Figs. \ref{fig:wafer-trade-off} (a–c) depicts three WSC configurations with varying DRAM chiplet counts. \textbf{Wafer\,0} places six DRAM chiplets on both sides of the compute die, maximizing memory capacity for storing large models and checkpoints. \textbf{Wafer\,1} reduces DRAM to three chiplets on one side, reallocating area to enlarge the compute die and improve both compute capability and D2D bandwidth. \textbf{Wafer\,2} retains the Wafer\,1's compute-die size but further reduces DRAM, freeing more peripheral IO to enhance D2D bandwidth. Fig. \ref{fig:wafer-trade-off} (d) summarizes the inherent architecture trade-off. \textbf{(1)} Increasing DRAM capacity reduces space available for compute dies, in turn lowering compute density. \textbf{(2)} Given the fixed number of compute-die interconnect interfaces, allocating more interfaces to DRAM reduces the bandwidth available for D2D links.

\subsection{Untapped Potential of Training Strategy $\&$ WSC Co-design}
Fig. \ref{fig:co-design} (a) illustrates three checkpoint (i.e., recomputation) strategies (Type\,0-2) applied to a computation graph with three operators (L1-L3) with TP=2, where each operator represents the corresponding operation across all layers. Type\,0 stores all activation checkpoints without recomputation. Type\,1 recomputes outputs of operator L2, while Type\,2 recomputes L1's output. Due to operator-specific tensor partitioning, each strategy incurs distinct compute, storage, and communication demands. As depicted in Fig. \ref{fig:co-design} (b), Type\,0 has the highest storage cost but minimal computation overhead, as it avoids recomputation entirely. In contrast, Type\,1 and Type\,2 incur higher compute overheads, but differ in storage and communication footprints based on which activations are recomputed. 

These distinct training strategies impose specific demands for on-wafer resource allocation, underscoring the co-design of training strategies and architecture. As depicted in Fig. \ref{fig:co-design}(c), traditional designs often focus solely on strategy-level optimizations, while neglecting architectural adaptation, leading to sub-optimal real performance. By contrast, co-design leverages dominant tensor partitioning and recomputation patterns to proactively tailor wafer architectures to the needs of typical training workloads, seeking enhanced overall performance.

Unfortunately, existing DSE frameworks almost fail to leverage the co-optimization of training strategies and architecture, let alone account for wafer-level architectural constraints. Table \ref{tab:works_comparision} summarizes their features. Timeloop \cite{parashar2019timeloop} focuses on die-level mapping exploration and architecture optimization, while Hecaton \cite{huang2024hecaton} and Gemini extend to the chiplet scale. DFModel \cite{ko2024dfmodel}, Calculon \cite{isaev2023calculon} and Bpipe \cite{kim2023bpipe} explore optimal training strategy or scheduling on GPU clusters, but none of these works consider the unique constraints of WSCs. Recent works like WSC-LLM \cite{xu2025wsc} and PD \cite{yang2025pd} realize WSC constraints, but one focuses on inference and the other on topology optimization, lacking a comprehensive training strategy. They overlook custom training optimizations for recomputation and checkpoint overheads, which motivates the development of WATOS.

% Consequently, to alleviate memory pressure at early pipeline stages during LLM training on WSCs, substantial recomputation is required. However, this introduces additional overhead and severe pipeline bubbles. As shown in Fig. \ref{fig:pipeline}, the overall pipeline iteration time is dictated by the stage with the highest recomputation cost. Therefore, balancing memory utilization across stages is essential to minimize recomputation and reduce training latency. 

% Consequently, significant recomputation is incurred to alleviate memory pressure at early stages when training large models on WSCs, yet it will lead to extra overhead and severe pipeline bubbles \cite{fan2021dapple}. As shown in Fig. \ref{fig:pipeline}, the pipeline iteration time depends on the stage with maximal recomputation overhead. Hence, balancing memory utilization is essential for reducing recomputation and eventually reducing training iteration time.

\section{WATOS Framework}\label{sec:WATOS_framework}

\begin{figure*}[t]
\centering
\includegraphics[width=0.97\linewidth]{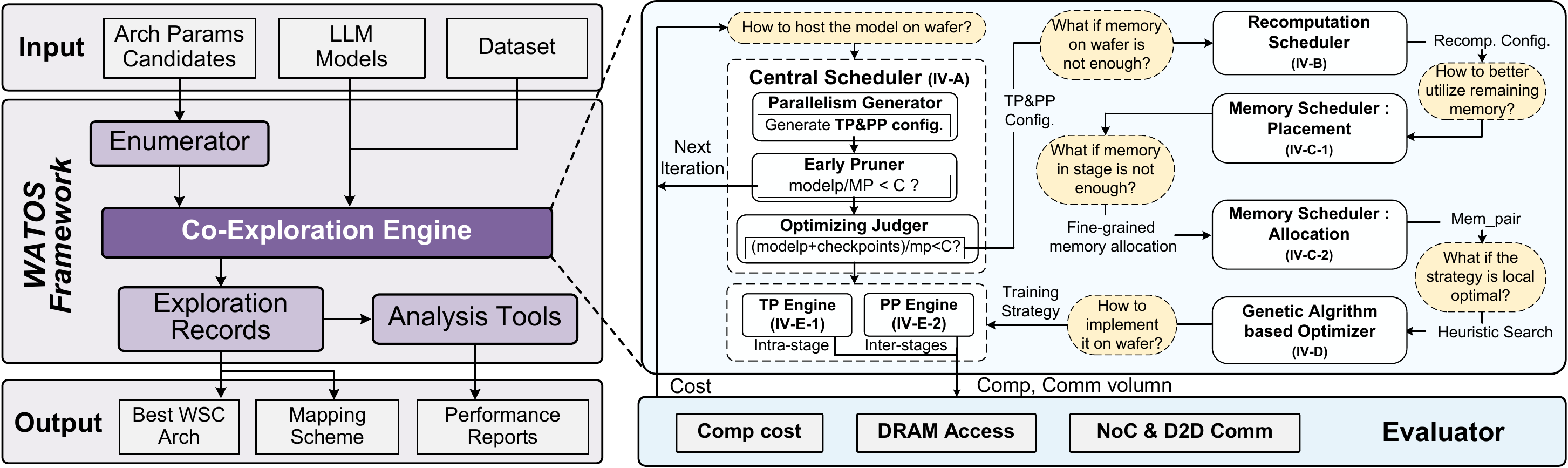}\vspace{-2mm}
\caption{The proposed WATOS framework for LLM training and wafer-scale architecture co-exploration.}
\label{fig:WATOS_framework}\vspace{-2mm}
\end{figure*}

As depicted in the left of Fig. \ref{fig:WATOS_framework}, WATOS takes architecture parameter candidates, LLM model configs, and training datasets as inputs. Considering wafer area constraints, architecture parameter candidates are derived from arbitrary combinations of configurable parameters (as in \S\ref{subsec:wafer-scale-chip}). All architectural candidates are first exhaustively explored in an \textit{Enumerator}, then sent to the \textit{Co-exploration Engine} to determine the achievable optimal LLM training performance on WSCs.

\textbf{Overall workflow}. As shown in the right part of Fig. \ref{fig:WATOS_framework}, the central scheduler first iterates through feasible TP and PP sizes, and preliminarily allocates resources across pipeline stages. Then, the recomputation scheduler prevents memory overflow by selectively recomputing activations with minimal latency overhead. Next, the memory scheduler optimizes on-wafer memory usage by topology-aware resource placement and fine-grained DRAM allocation. Further, the global optimizer applies genetic algorithms to search the remaining design space and identify optimal configurations. Finally, the execution engines generate detailed implementation plans, invoke the \textit{Evaluator} to assess performance, and return results to the central scheduler for the next exploration round. 

\subsection{Central Scheduler}\label{subsec:central_scheduler}
%我觉得Cental Scheuler的作用是先依据Workload筛选架构，再在架构上进行资源分配推到下一个模块，并记录架构的最佳策略
%Given the \emph{DRAM penalty} on WSCs, we prioritize model parallelism within a single wafer, which requires careful partitioning of TP/PP resources. However, this raises two key challenges: determining the appropriate TP/PP ratio, which must account for physical topology characteristics of WSCs. Second, the diversity of models and datasets imposes different requirements on WSC architectures. It is therefore critical to identify the architectural feasibility in order to enable further and targeted computation and memory optimization.

%Given the DRAM penalty on WSCs, we prioritize model parallelism, which necessitates careful TP and PP partitioning. To tackle this challenges, we propose Alg. \ref{alg:resoure_allocation}, which systematically optimizes the resource partitioning for TP and PP.

Given the trade-off between memory, computation, and communication on WSCs, we prioritize model parallelism within a single wafer, requiring careful TP and PP partitioning. This presents two challenges: determining the optimal TP/PP ratio while considering architectural constraints, and managing activation checkpoints to prevent out-of-memory (OOM) errors

% Given the trade-off between memory, computation, and communication on WSCs, we prioritize model parallelism within a single wafer, which requires careful partitioning of TP and PP resources. However, this raises two key challenges: determining the appropriate TP/PP ratio, which must account for the architectural constraints. Second, the explosive activation checkpoints in training process may lead to out-of-memory (OOM) errors. It is therefore critical to detect potential OOM errors and determine whether to integrate strategies such as recomputation to prevent them.

To this end, we propose Alg. \ref{alg:resoure_allocation}, which systematically optimizes resource partitioning for TP and PP. Our optimization is based on the  premise that training data consists of four parts: model weights, gradients, optimizer states, and activation checkpoints. The first three, collectively referred to as \emph{modelP}, are essential and must be retained throughout the training process, while activation checkpoints are optional and can be regenerated through recomputation. Thus, minimal model-parallel units must fit \emph{modelP} in their aggregate memory.

Given a model $M$, model-parallel dies $MP$, per-die DRAM capacity $C$, workload $W$, the central scheduler first checks if \emph{modelP} fits within the max aggregate memory capacity of the MP. If not, it will directly prune the parallelism candidate (line 1, 2). It then iterates over all feasible TP and PP sizes, for each $mp$ within the $MP$ (line 4), ensuring the TP instance consists of an even number of dies to meet the 2D-mesh topology communication requirements. Once the TP and PP are determined, the algorithm checks if the checkpoint storage requirement is met (line 5). For failed configurations, it delegates them to downstream schedulers, which employ recomputation and memory allocation to ensure compliance (line 6). Finally, the algorithm enumerates all possible TP partition strategies (line 7, detailed in \S\ref{subsec:TP_engine}) to identify the max achievable throughput for each configuration, updating the current configuration if a better record is found (line 8).

\begin{table}[t]
\renewcommand{\arraystretch}{1.05}
\caption{Summary of Frameworks (Co-de: Co-design, Opt: Optimization)}\vspace{-2mm}
\centering
\begin{threeparttable}
\footnotesize
\begin{tabular}{l||m{0.5cm}<{\centering}|m{0.45cm}<{\centering}|m{0.4cm}<{\centering}|m{0.75cm}<{\centering}|m{0.7cm}<{\centering}|m{0.6cm}<{\centering}|m{0.45cm}<{\centering}}
\specialrule{0.12em}{0.5pt}{0.4pt}

 \multirow{2}{*}{\!\!\!\textbf{Framework}} &   \multicolumn{3}{c|}{\textbf{HW Optimization}} &  \!\!\!\textbf{Recomp}\!\!\!\! & \textbf{WSC} & \!\!\textbf{WSC}\!\! & \!\!\textbf{Opt}\!\! \\
\cline{2-4}
& \!\!\!\!\! {Comp}\!\!\!  & \!\!\! {Mem}\!\! & \!\!\! {D2D}\!\!   &  \!\!\!\textbf{aware}\!\! & \!\!\!\textbf{Physical}\!\!\! & \!\!\textbf{Co-de}\!\! & \!\!\!\textbf{level}\!\!\!\\
\hline
% \rowcolor{mygray}\!\!\!$\mathbf{A}^3$\! \cite{ham20203} & $\times$ & \checked  & $\times$ & $\times$ & P only & Value\\
\!\!\!\textbf{Timeloop} \cite{parashar2019timeloop} & \checked & $\times$ & $\times$ & $\times$ & $\times$ & $\times$ & \!\!die \\
\rowcolor{mygray}\!\!\!\textbf{Hecaton} \cite{huang2024hecaton}  & \checked  & \checked & \checked & $\times$ & $\times$ & $\times$ & \!\!chiplet\\
\!\!\!\textbf{Gemini} \cite{cai2024gemini} & \checked & \checked & \checked & $\times$ & $\times$ & $\times$ & \!\!chiplet\\
\rowcolor{mygray}\!\!\!\textbf{DFModel} \cite{ko2024dfmodel} & \checked   & $\times$ & $\times$ & $\times$ & $\times$ & $\times$ & \!\!cluster\\
\!\!\!\textbf{Calculon} \cite{isaev2023calculon} & \checked & \checked & $\times$ & \checked & $\times$ & $\times$ & \!\!cluster\\
\rowcolor{mygray}\!\!\!\textbf{BPipe} \cite{kim2023bpipe}\!\!\! &  \checked   & \checked & \checked & $\times$ & $\times$ & $\times$ & \!\!cluster\\
\!\!\!\textbf{FRED} \cite{rashidi2024fred}  & \checked  & $\times$ & \checked & $\times$ & $\times$ & \checked & \!\!wafer\\
\rowcolor{mygray}\!\!\!\textbf{PD} \cite{yang2025pd}\!\!\!\!\! & \checked   & $\times$ & \checked & $\times$ & \checked & \checked & \!\!wafer\\
\!\!\!\textbf{WSC-LLM}\cite{xu2025wsc}\!\!\!\!\! & Low   & $\times$ & $\times$ & $\times$ & \checked & \checked & \!\!wafer\\
\rowcolor{mygray}\!\!\!\textbf{WATOS} & {\fontsize{6.5}{6} \CheckmarkBold }  & {\fontsize{6.5}{6} \CheckmarkBold } & {\fontsize{6.5}{6} \CheckmarkBold }  & {\fontsize{6.5}{6} \CheckmarkBold } & \fontsize{6.5}{6} \CheckmarkBold & \fontsize{6.5}{6} \CheckmarkBold & \!\!\textbf{wafer}\\
\specialrule{0.12em}{0.5pt}{0.1pt}
\end{tabular}
% \begin{tablenotes}
% \footnotesize
% \item \hspace{-5mm}*  \!\!\! Sparsity guided by preceding layer scores; Accuracy degradation w/o retrain.
% \end{tablenotes}
\end{threeparttable}
\label{tab:works_comparision}
\end{table}

\begin{figure}[t]
\centering
\includegraphics[width=\linewidth]{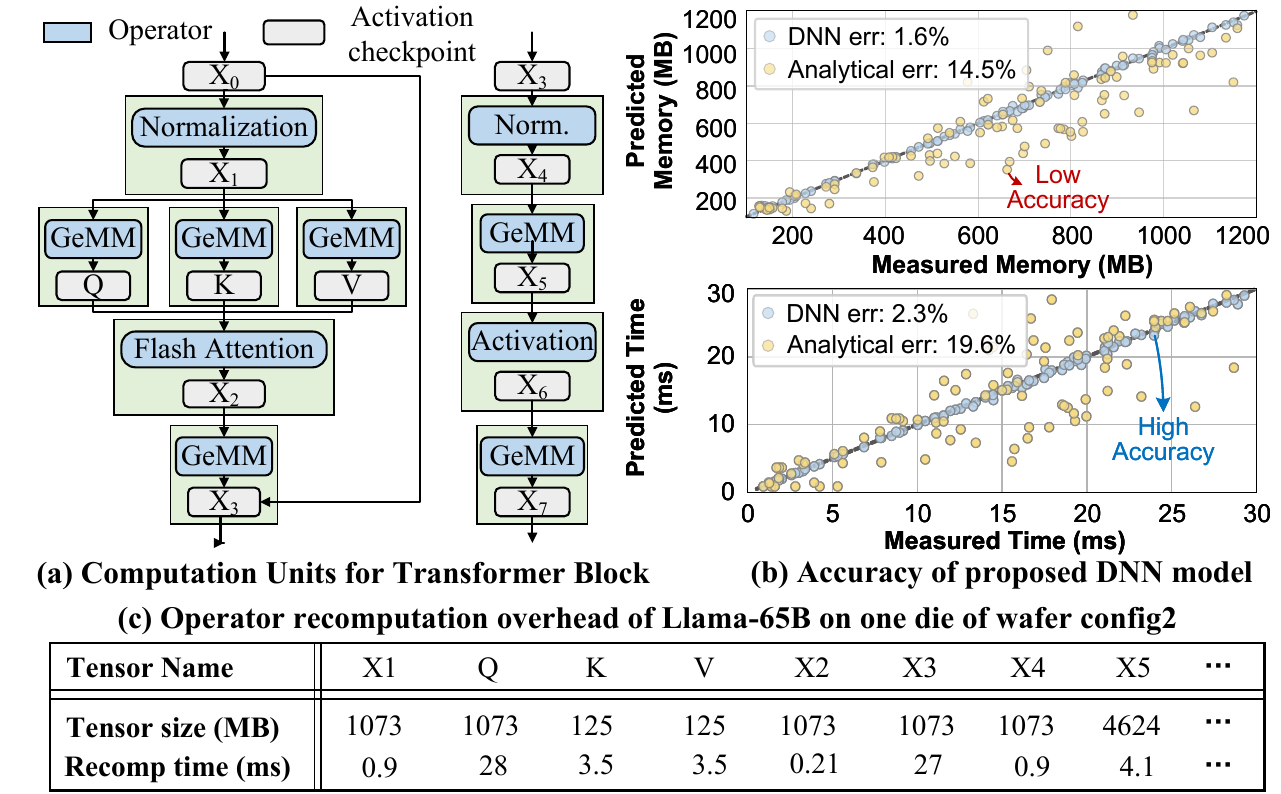}\vspace{-2mm}
\caption{Recomputation preparation for LLM models.}
\label{fig:operator}
\end{figure}

\subsection{Recomputation Scheduler}\label{subsec:computation_scheduler}

% In \S \ref{subsec:central_scheduler}, we derive the parallel configuration (TP\_Config, PP\_Config) that fits the \emph{modelp} with partial activation checkpoints. However, remaining checkpoints exceeding on-wafer capacity may cause OOM errors. Recomputing these checkpoints alleviates memory constraint but naive strategies introduce bubbles and underutilize memory. To this end, we propose a globally coordinated memory-efficient recomputation (GCMR) strategy (Alg. \ref{alg:Recomputation_strategy}), which balances recomputation to minimize bubbles and maximize utilization. To address early-stage overflows in 1F1B pipeline, GCMR identifies memory-constrained stages as Senders and memory-rich stages as Helpers, offloading checkpoints to balance memory usage.

In \S \ref{subsec:central_scheduler}, we obtain the parallel configuration (TP$\_$Config, PP$\_$Config) that can accommodate the \emph{modelp} with partial activation checkpoints. The remaining checkpoints exceeding on-wafer memory capacity can result in OOM errors that interrupt the training process. To address this limitation, recomputing these checkpoints enables training to proceed within the given memory constraints. However, as shown in Fig.~\ref{fig:pipeline} (b)(c), the naive recomputation strategy not only introduces pipeline bubbles but also underutilizes memory, thereby incurring additional computation overhead and degraded training throughput.

To this end, we propose a globally coordinated memory-efficient recomputation (GCMR) strategy, as outlined in Alg. \ref{alg:Recomputation_strategy}. In contrast to the naive recomputation strategy depicted in Fig. \ref{fig:pipeline} (a), the GCMR strategy takes into account the DRAM capacity across the entire WSC. It employs a global balanced recomputation strategy that minimizes the bubble while maximizing memory utilization, as demonstrated in Fig. \ref{fig:pipeline} (b). Notably, the memory imbalance introduced by 1F1B pipeline may still lead to checkpoint overflows in early stages, like the stage0 and stage1 in Fig. \ref{fig:pipeline} (c). Therefore, after generating the recomputation configuration \emph{Recomp\_config}, the GCMR identifies stages that are susceptible to such imbalance as Senders and those with sufficient memory as Helpers. The Helpers can receive the overflowing \emph{checkpoints} from the Senders, thus balancing memory usage across the pipeline.

Given that most LLMs are based on the Transformer architecture, we begin with decomposing the computation into a set of fundamental operator units according to the Transformer graph, as shown in Fig. \ref{fig:operator} (a). Each operator is annotated with its computation type as well as the associated checkpoint requirements, enabling fine-grained recomputation scheduling. We further incorporate the SOTA FlashAttention \cite{dao2023flashattention} mechanism as a specialized operator to capture its distinct performance and memory characteristics. To cover all possible setups, we train a DNN model to predict the execution latency and memory footprint of each operator, based on two key considerations: \textbf{1)} Despite their precision, cycle-accurate simulators often require minutes to hours per run, making them impractical for large-scale performance evaluation. \textbf{2)} Analytical models, on the other hand, fail to capture complex factors such as alignment overheads and multi-level memory effects, resulting in limited prediction accuracy. As depicted in Fig. \ref{fig:operator} (b), DNN-based prediction can realize up to $17.3\%$ higher accuracy over the analytical scheme. \textbf{3)} Numerous simulator works have validated the feasibility of using DNNs, with prediction error remaining within a controllable range \cite{panner2024can, kim2024llmem, xu2025wsc}. Therefore, we adopt a DNN model to predict the execution latency and memory footprint of each operator under different batch sizes and wafer-scale hardware configurations. The collective communication overhead of recomputed tensors between adjacent operators is estimated through the $\alpha$–$\beta$ model \cite{thakur2005optimization} as Eq. \eqref{eq:alphabeta}, where $\alpha$ represents link latency and $\beta$ denotes the communication size of All-Reduce. This profiling is conducted offline and produces an operator-level performance lookup table for iterative process, as depicted in Fig. \ref{fig:operator} (c).
\begin{equation}
\label{eq:alphabeta}
{t_{comm}=\alpha+{\beta}/{BW},\ \ \  \ \beta=2\cdot{(TP-1)}/{TP}\cdot BSH}
\end{equation}

\begin{algorithm}[t]
\begingroup
\linespread{1.0}\selectfont
\small
\setlength{\baselineskip}{0.96\baselineskip}
\caption{\small{Early Pruning-based Central Scheduler}}
\label{alg:resoure_allocation}
\small
\KwIn{
    Model $M$, die limit for MP $MP$, Die memory capacity $C$, Workload $W$, TP split strategy $S$
}
\KwOut{OptimalConfig}
\If{$W.modelP/MP > C$}{
    Prune(Candidate)\;
}
\Else{
    \For{$tp\cdot pp == mp \in [1, MP]$}{
        \If{$(W.modelP + W.checkpoints) / mp > C$}{
            DownstreamSchedulers(tp, pp)
        }
        \For{$s \in S$}{
            OptimalConfig $\gets$ $\mathbf{argmax}(test(W,tp,pp,s))$
        }
    }
}

% \Return{OptimalConfig}
\endgroup
\end{algorithm}

Based on the computation and memory characteristics of fundamental LLM operators, GCMR performs dynamic programming to accelerate the search for the optimal recomputation strategy. It begins from the last pipeline stage, incrementally allocating memory and querying the profile for latency and recomputation candidates, recursively incorporates preceding stages. After analyzing all stages, the algorithm identifies the optimal configuration by minimizing the maximum stage-execution time (line\,2–5). Finally, the recomputation configurations along with memory footprints for each stage are generated (line\,6-8). In this way, the GCMR achieves maximal global memory efficiency at minimal recomputation cost.

Moreover, lines 9–11 identify Senders and Helpers by examining the post-recompute memory footprint of each stage. Specifically, stages exceeding per-die memory capacity are marked as \emph{Sender} for \emph{checkpoints} offloading (lines\,8,9), while others serve as \emph{Helper} to receive the checkpoints (line\,10,11). The algorithm then sorts the Sender and Helper sets by memory pressure and available memory (line 12) and invokes the Mem\_pair procedure to pair memory-constrained Senders with memory-rich Helpers (lines 13,14). This pairing offloads the activation checkpoint from high-pressure stages to memory-rich stages, effectively balancing memory usage across the pipeline. \textbf{Notably, the offloading occurs across dies on a wafer, rather than off-wafer to external memory.}

\begin{algorithm}[t]
\begingroup
\linespread{0.99}\selectfont
\small
\setlength{\baselineskip}{0.95\baselineskip}
\caption{\small{The Proposed GCMR Strategy}}
\label{alg:Recomputation_strategy}
\KwIn{TP Config, PP Config, Workload \textbf{W}}
\KwOut{RecompConfig $\mathcal{R}$, SenderSet $\mathcal{S}$, HelperSet $\mathcal{H}$}
P$\gets$ RecompProfiling(pp, tp, W)\;
\For{$t = pp-1:0$}{
\For{$m = 0:(pp-t)*C$}{
    $t_{max} \gets max(T[t+1,m-m_t],B_t+2F_t-P(m_t))$\;
    $T[t,m] \gets \mathop{min}\limits_{m_t}(t_{max}),\!$ Mem[t,m]$\gets\! \mathop{argmin}\limits_{m_t}(t_{max})$\;
}
}
\For{$t = 0: pp-1$}{
    M[t]$\gets$Mem[$t,pp\cdot C-\sum^{t-1}_{i=0}M[i]$], $\mathcal{R}$[t]$\gets$ P(M[t])\;
    
    \eIf{$M[t]>C$}{
        $\mathcal{S}.append(stage_t)$\;
    }{
        $\mathcal{H}.append(stage_t)$\;
    }
}

$\mathcal{S}, \mathcal{H} \gets DescendSort(\mathcal{S}, \mathcal{H})$\;
\While{$\mathcal{H\neq \varnothing }$}{
    Mem\_pair($\mathcal{S}$.pop(), $\mathcal{H}$.pop())\;
}
%\Return{$\mathcal{R}$, $\mathcal{S}$, $\mathcal{H}$, Mem\_pair}
\endgroup
\end{algorithm}

\subsection{Memory Scheduler}\label{subsec:memory_scheduler}
\textbf{1. Optimal Resource Placement Strategy}.\label{subsubsec:MemoryPlacement} In \S~\ref{subsec:computation_scheduler}, we investigate the memory-efficient recomputation configurations, along with the construction of the Mem\_pair set. This pairing enables offloading checkpoints from the Sender to the DRAM chiplets of the Helper. However, due to characteristics of the 2D-mesh topology on wafer-scale chips, it is also necessary to determine the optimal placement of each instance.

%Each Mem\_pair consists of a pair of stages: a Sender stage experiencing memory pressure beyond the DRAM capacity, and a Helper stage with available DRAM headroom.

%\begin{figure*}
% \vspace{-2mm}
% \begin{equation}
% \small
%     GlobalCost = min \left(\sum_{i=0}^{pp-1} Dist(S_i,S_{i+1})\cdot Comm_{pp}(S_i,S_{i+1})\right. \left. \\
%     + \!\!\!\!\!\!\!\!\!\sum_{(s,h)\in Mem\_pairs}\!\!\!\!\!\!\!\!\!\!Dist(S_s, S_h)\cdot Comm_{pair}(S_s,S_h)\cdot(1+\gamma)\right).
% \label{eq:global_cost}
% \end{equation}
%\end{figure*}

\begin{figure}[t]
\includegraphics[width=0.94\linewidth]{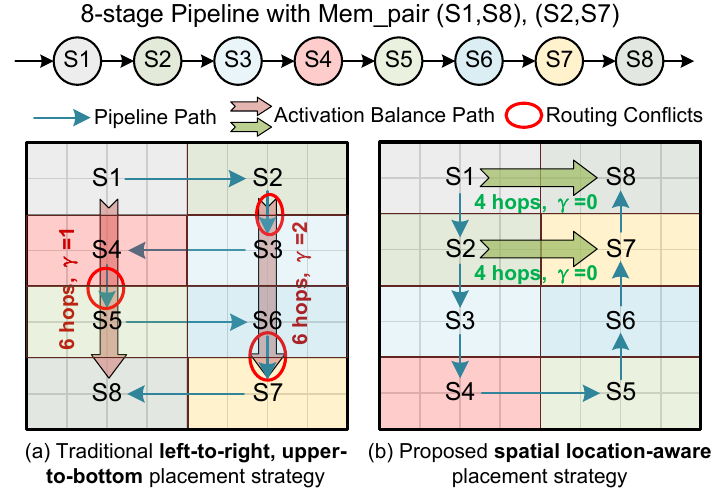}\vspace{-2mm}
\caption{Comparisons of different resource placement strategies.}
\label{fig:placement}
\end{figure}

Fig. \ref{fig:placement} illustrates two resource placement strategies for an 8 stage pipeline with designated Mem\_pair configurations $(S1,S8)$ and $(S2,S7)$. The pipeline stages are arranged linearly from S1 to S8, with activation checkpoint balance paths identified between the corresponding Mem\_pairs. As can be seen, the conventional left-to-right, upper-to-bottom resource placement strategy results in long communication paths between Mem\_pair stages, with an average of 6 hops. In contrast, our proposed spatial location-aware placement approach strategically co-locates Mem\_pair stages, significantly reducing the communication cost by shortening the path length from 6 hops to 4 hops. Moreover, when considering the overall communication hops, including those along the pipeline path, our strategy achieves a 30\% reduction in total hop count.

Since the activation checkpoint transfer occurs from Sender instances to Helper instances, our objective is to minimize the total distance between each Sender instance and its Helper instance, while avoiding overlapping transmission paths with the pipeline paths. To achieve this, we adopt a globally location-aware placement strategy that preserves the critical path of the pipeline, while minimizing the transmission distance between Mem\_pair stages. If multiple placement schemes exist, we select the one that minimizes the total communication cost. The selection is formalized as Eq. \eqref{eq:globalcost}, 

\vspace{-4mm}
\begin{equation}
\label{eq:globalcost}{ 
\small
\begin{aligned}
    \small
    GlobalCost = min \left(\sum_{i=0}^{PP-1} Dist(S_i,S_{i+1})\cdot Comm_{PP}(S_i,S_{i+1})\right. \\
     \!\!\!\!\!\!\!\!  +  \!\!\!\!\!\!\!\!\!\!\!\!\!\!\!\!\!\sum_{(s,h)\in Mem\_pairs}\!\!\!\!\!\!\!\!\!\!\!\!\!\left.Dist(S_s, S_h)\cdot Comm_{pair}(S_s,S_h)\cdot(1+\gamma)\right)\!,
\end{aligned}}
\end{equation}

\noindent where $S_i$ denotes the center position of the pipeline stage, $i\in [0,PP-1]$.  $Comm_{PP}$ and $Comm_{pair}$ represent the pipeline and checkpoint transfer communication workload respectively, which are used as weights for distance. $Dist$ refers to the shortest path between the two pipeline stages. In cases where multiple shortest paths exist, all such paths are enumerated. To take routing conflicts into account of placement, the conflict factor $\gamma$ is introduced, which stands for the number of conflict links between activation balance paths and existing pipeline paths. Paths with conflicts will be punished by $1+\gamma$ in $GlobalCost$.

\noindent \textbf{2. Location-aware DRAM capacity allocation}. WSCs feature high D2D bandwidth, typically exceeding DRAM access bandwidth. This means that, cross-die DRAM read and write operations are limited by DRAM bandwidth rather than D2D bandwidth. Leveraging this advantage of the wafer-scale architecture, the extensive inter-die transfer of checkpoints can be overlapped by its DRAM access, allowing us to fully utilize all DRAM resources on the wafer for the activation checkpoint storage, to achieve higher performance of LLM training.

%Based on GlobalCost, we implement a location-aware DRAM allocation strategy to determine the optimal pairing of \emph{Mem\_pair}, aiming to minimize the D2D transmission overhead, as depicted in Algorithm \ref{alg:DRAM_allocation}. Given the Sender set $\mathcal{S}$ and the Helper set $\mathcal{H}$, the algorithm first computes and sorts the GlobalCost (Eq. \ref{eq:globalcost}) of all Sender–Helper pairs, forming a priority queue $Q$ (line 2). For instance, in placement strategy as Fig. \ref{fig:placement}(b), the DRAMs on $S_2$, $S_8$, and $S_7$ occupy top positions in the priority queue $Q$ of $S_1$ for the short transmission paths between their respective positions. Then, we employ a greedy heuristic to avoid the NP-hard problem arising from checkpoint capacity constraints and heterogeneous communication costs. For each $S_i$, the algorithm iteratively selects the Helper with the lowest cost (lines 5–6). If the Helper can store the current checkpoint, it remains available for other Senders (lines\;7–9); otherwise, the Sender proceeds to the next best Helper (lines\;10–11). The allocation loop continues until the entire checkpoint requirement of $S_i$ is met.

As depicted in Alg. \ref{alg:DRAM_allocation}, we implement an location-aware DRAM allocation strategy for handling overflow checkpoints, to minimize the additional D2D transmission overhead. Given the Sender set $\mathcal{S}$, the Helper set $\mathcal{H}$, the algorithm finds the optimal DRAM destination for each Sender's checkpoint. The algorithm first identifies the associated Helpers for each Sender without additional communication cost to form a priority queue $Q$, based on the existence of short transmission paths between their respective positions (line 2). For example, in Fig.10 (b), DRAMs within S2, S8 and S7 occupy top positions in the queue $Q$ of $S_1$. Subsequently, an empty Mem\_pair $A_i$ is initialized (line 3), and the algorithm iteratively allocates DRAM resources to the Sender $S_i$. The capacity of DRAMs of Helper is reduced by the portion utilized during the allocation process (line 5-7). After calculating the remaining capacity, the DRAM unit is reinserted into the queue Q (line 7-9), which is dynamically reprioritized to guide subsequent allocations. The allocation loop continues until the entire checkpoint requirement of $S_i$ is met. Notably, the Alg. \ref{alg:DRAM_allocation} does not conflict with the optimization in \ref{subsec:memory_scheduler}-1, but rather builds upon it to execute a fine-grained DRAM allocation strategy.

\begin{algorithm}[t]
\begingroup
\setlength{\baselineskip}{1.0\baselineskip}
\small
\caption{\small{Location-aware DRAM Allocation Strategy}}
\label{alg:DRAM_allocation}
\KwIn{Sender set $\mathcal{S}$, Helper set $\mathcal{H}$\!
}
\KwOut{Memory pair set $\mathcal{A}=\{A_1,..,A_k\}$}
\For{$S_i \in \mathcal{S}$}{
   $Q\gets $ sort$(GlobalCost(S_i, \mathcal{H}))$\;
   $A_i\gets \varnothing$\;
   \While{$S_i.OverflowMemory>0$}{
   $d'\gets Q.pop()$;~$A_i.append(d')$\;
   \If{$d'.capacity > S_i.OverflowMemory$}{
   $d'.capacity$ $-\!=$ $S_i.OverflowMemory$\;
   $Q.insert(d')$\;
   }
   \Else{
      $S_i.OverflowMemory$ $-\!=$ $d'.capacity$\;
   }
   }
}
%\Return{$\mathcal{A}$}
\endgroup
\end{algorithm}

\subsection{Global Optimizer}\label{subsec:GlobalOptimizer}
\begin{figure}[t]
\centering
\includegraphics[width=0.95\linewidth]{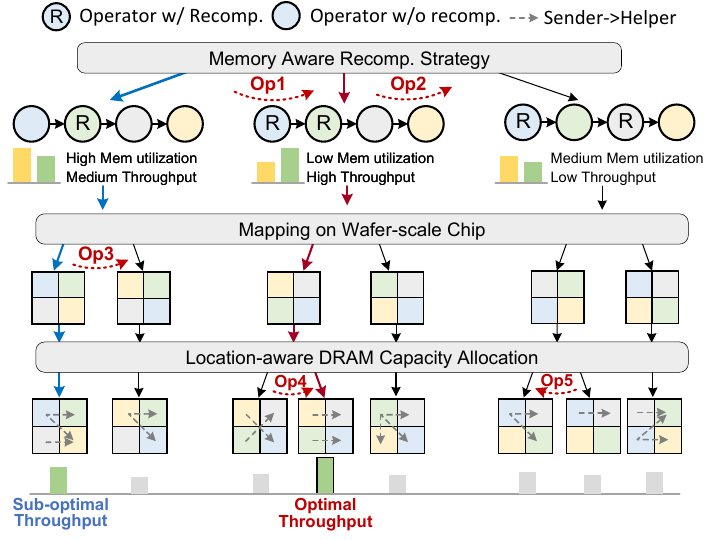}\vspace{-2mm}
\caption{Overview of genetic algorithm-based global optimizer.}
\label{fig:GA}
\end{figure}

\S\ref{subsec:computation_scheduler} and \S\ref{subsec:memory_scheduler} present the optimal recomputation configuration and DRAM allocation strategy, respectively. However, maximizing memory utilization alone cannot guarantee full overlap of recomputation bubbles. The greedy strategy may fail to pair each Sender with the lowest-cost Helper, leading to local optima, as blue path in Fig. \ref{fig:GA}. To overcome this limitation, we employ a genetic algorithm (GA) \cite{holland1992genetic}, which can identify global optima within vast solution spaces based on evolutionary principles \cite{schmitt2001theory}. 

By integrating GA with customized operators, WATOS can thoroughly explore the design space toward near-optimal solutions. Specifically, given Recomp\_config $\mathcal{R}$, Mem\_pair $\mathcal{A}$ and placement strategy, we develop five GA operators to facilitate the exploration process, as Op1-Op5 in Fig. \ref{fig:GA}:

\textbf{Op1 $\mathcal{R}$ variation}: Randomly enable or disable recomputation for an operator while satisfying DRAM capacity.

\textbf{Op2 $\mathcal{R}$ crossover}: Swap recomputation configurations between two randomly selected pipeline stages beyond a chosen backward operator.

\textbf{Op3 Placement variation}: Swap the physical positions of two randomly chosen stages on the wafer-scale chip.

\textbf{Op4 $\mathcal{A}$ variation}: Modify the Mem\_pair of a Sender by adding the lowest-cost Helper or removing part of a randomly selected Helper, while fulfilling the storage constraints.

\textbf{Op5 $\mathcal{A}$ crossover}: Exchange Mem\_pairs between two randomly selected Senders.

% With these operators, mutations can be introduced during each iteration, generating new configurations. Then a fitness function, calculated as the product of the longest stage execution time $t_{max}$ and \emph{GlobalCost}, evaluates the performance of all configurations. Subsequently, all configurations are probabilistically eliminated based on the fitness function to maintain tractable search. After multiple iterations, the configuration with the highest fitness is selected as the final solution.

With these operators, mutations and crossover are introduced to generate new configurations. Then, a fitness function, defined as $t_{max}\times GlobalCost$, evaluates performance, and configurations are probabilistically eliminated to keep the search tractable. Crucially, these operators enable any feasible configuration to evolve into another through valid transformations. For instance, as depicted in Fig. \ref{fig:GA}, Op1 can introduce additional recomputation into the initial recomp\_config, improving pipeline overlap at the expense of minor DRAM utilization. Then Op4 adjust Mem\_pairs obtained from the greedy strategy, sacrificing the local optimum to reach global optima as red path in Fig. \ref{fig:GA}. Therefore, the Global Optimizer ensures that any point within the design space of WSC can be reached from any other point, thereby guaranteeing comprehensive exploration and yielding global optimal solutions.

%For instance, the initial recomp\_config can add a recomputation operator, slightly reducing DRAM utilization to achieve a better overlapped pipeline through Op1. After the placement, Op4 can be applied to exchange the Mem\_pair obtained from the greedy strategy, sacrificing the local optimum for a single Sender to achieve a global optimum. 

%With these operators, the GA 每次迭代在上个周期的种群进行变异得到更多的个体。随后计算所有个体的适应度函数，基于此值概率淘汰个体以控制种群规模。Global Optimizer以最长stage执行时间$t_{max}$和GlobalCost的乘积作为适应度函数，并以Computation scheduler和Memory Scheduler产生的解作为初始种群，完成迭代后选择适应度最高的个体作为最终配置。The GA ensures that any point within the LP SPM optimization space can be reached from any other, thereby guaranteeing comprehensive exploration and yielding near-optimal solutions.

%Crucially, this ensures that any point within the LP SPM optimization space can be reached from any other (demonstration link [1]), thereby guaranteeing comprehensive exploration by the SA algorithm, and yielding near-optimal solutions.Through our SA-based algorithm combined with our specially designed operators, Gemini not only can explore the optimization space to balance the trade-offs introduced in Sec IV-C, but also automatically optimize D2D link communication.

% \begin{figure}[t]
% \centering
% \includegraphics[width=0.87\linewidth]{figures/inter-stage3.pdf}\vspace{-3mm}
% \caption{Inter-stage communication mechanism.}
% \label{fig:inter-stage}\vspace{-1mm}
% \end{figure}

\subsection{TP \& PP Execution Engines}\label{subsec:TP_engine}
% With optimal LLM training strategy, TP $\&$ PP Execution Engine 取得物理实现方案。
%To enable efficient execution of LLM operators on wafer, computational workloads must be carefully partitioned and mapped across dies and cores. As shown in Fig. \ref{fig:TP_engine}, a GEMM operator can be partitioned along four dimensions: batch size (B), sequence length (S), hidden size (H), and the reduction dimension (K). WATOS adopts a two-level computational task allocation strategy. (1) Die-level allocation (\S\ref{subsubsec:TP_engine}): The TP Engine decomposes the global computation into sub-computation tasks and distributes them across dies. (2) Core-level allocation (\S\ref{subsubsec:Core_engine}): Each sub-task is further partitioned into basic computation tiles, with each core executing one basic computation tile per cycle. 

% With an optimal LLM training strategy, the TP \& PP Execution Engines provide the physical implementation framework. 

%To efficiently execute LLM operators on wafers, workloads must be partitioned and mapped across dies and cores. As shown in Fig. \ref{fig:TP_engine}, a GEMM operator can be partitioned along four dimensions: batch size (B), sequence length (S), hidden size (H), and the reduction dimension (K). WATOS employs a two-level allocation strategy: (1) Die-level allocation (\S\ref{subsubsec:TP_engine}-1): where global computations are decomposed into sub-tasks and distributed across dies. (2) Core-level allocation (\S\ref{subsubsec:Core_engine}-2): where sub-tasks are further partitioned into computation tiles, with each core executing one tile per cycle.

\begin{figure}[t]
\centering
\includegraphics[width=\linewidth]{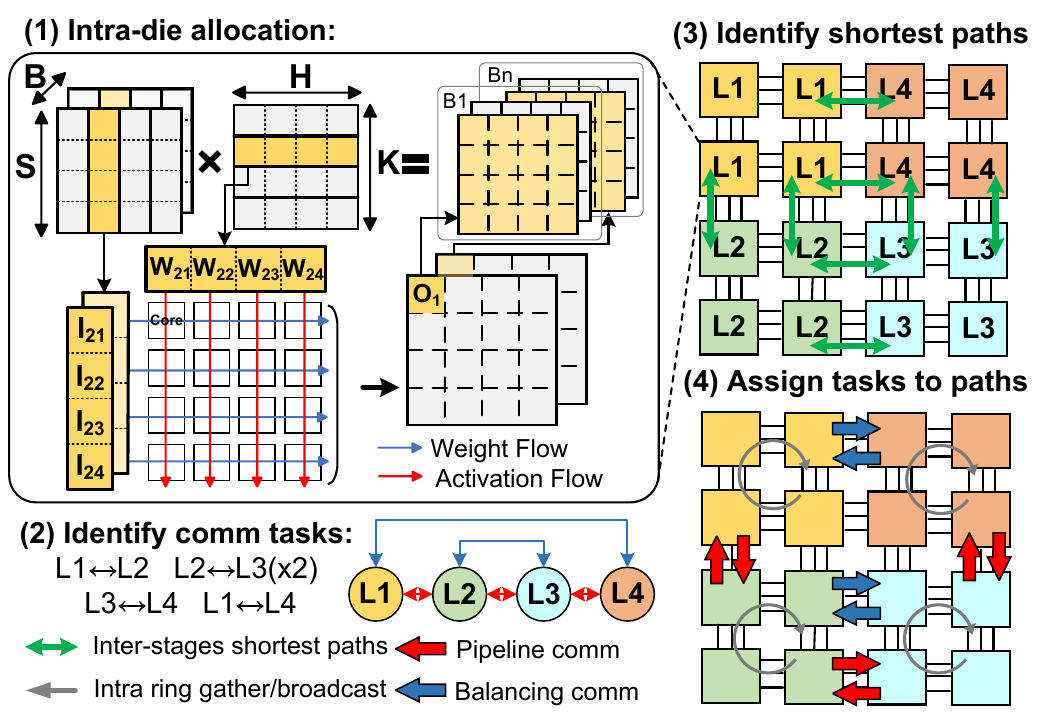}\vspace{-2mm}
\caption{Overview of TP\&PP execution engine.}
\label{fig:executionengine}
\end{figure}

To efficiently execute LLM operators on wafers, workloads must be partitioned and mapped across dies and cores. As shown in Fig. \ref{fig:executionengine}, a GEMM operator can be partitioned along four dimensions: batch size (B), sequence length (S), hidden size (H), and the reduction dimension (K). WATOS employs a two-level allocation strategy: Intra-stage allocation (\S\ref{subsubsec:TP_engine}-1) and Inter-stage allocation (\S\ref{subsubsec:PP_Engine}-2).

\begin{figure}[t]
\centering
\includegraphics[width=\linewidth]{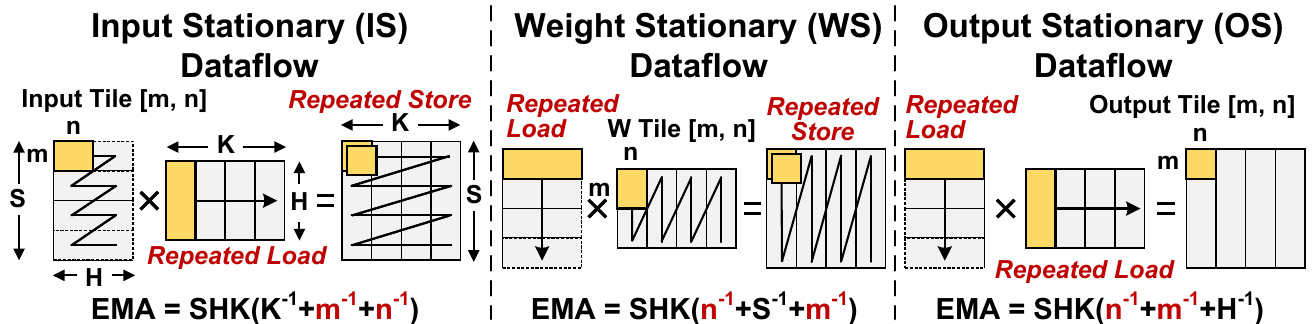}\vspace{-2mm}
\caption{Hybrid dataflows and EMA analysis using an $m\times n$ MAC array.}
\label{fig:hybriddataflow}
\end{figure}

\textbf{1. TP Engine.}\label{subsubsec:TP_engine} Initially, the TP engine decomposes computation into die-level sub-computation tasks. Once receiving a sub-computation task, each die partitions it into basic computation tiles (e.g., along B, S, H, K) sized to fit within a single core’s SRAM. These tiles are then mapped and executed. As depicted in Fig. \ref{fig:executionengine} (1), the TP Engine first partitions along the K dimension and assigns the sub-task to a die, which further partitions along S and H. Tiles are then mapped following an output-stationary (OS) dataflow and executed iteratively. However, OS may not fully maximize SRAM reuse, increasing system overhead. To address this, we further incorporate three additional dataflows: weight-stationary (WS), input-stationary (IS), and row-stationary (RS) \cite{chen2016eyeriss}. Among them, RS is specifically designed for convolution operators. Apart from RS, the variations of OS, WS, and IS do not change the computational workload but alter the external memory access (EMA), as illustrated in Fig. \mbox{\ref{fig:hybriddataflow}}, leading to differences in latency and energy consumption. Consequently, our intra-die dataflow adopts a hybrid design that dynamically selects the most suitable dataflow based on the operator type and EMA characteristics under different workloads. Upon completion of the computation, we implement all-gather and all-reduce operations in the TP process using the bidirectional ring algorithm \cite{zong2025ibing}, which aligns well with the 2D-mesh topology of WSCs. The TP engine can also be configured to explore other intra-stage communication mechanisms such as Multitree \cite{Huang2021Multitree}. Overall, the TP engine provides a robust approach to dynamically optimize intra-stage computation and communication, guaranteeing an efficient training process.

\textbf{2. PP Engine.}\label{subsubsec:PP_Engine} 
 Inter-stage communication is the major bottleneck in chiplet-based WSCs \cite{shao2019simba}. To mitigate contention from many-to-many communication, we adopt the following strategy for inter-stage communication, as shown in Fig. \ref{fig:executionengine}: Firstly, all inter-stage communication tasks, including pipeline data transfer between adjacent stages and activation balancing between senders and helpers, are identified. The shortest paths between corresponding stages are then determined. Subsequently, we allocate these communication tasks to links in order of size. In the allocation process, we prioritize unused links by giving a punishment factor to occupied links, and consequently avoid contention. As shown in Fig. \ref{fig:executionengine}, pipeline and activation balancing communications are overlapped as they occupy different links. Once the inter-stage communication cost determined, we can derive the execution time for a pipeline iteration based on the TP engine's execution strategy.

\subsection{Evaluator}\label{subsec:Evaluator}
The evaluator of the WATOS framework is built upon Astra-sim \cite{won2023astra}, an efficient and widely adopted simulator to model distributed parallel execution. Our \textit{Execution Engines} serves as the interface for global performance evaluation. To realize this, we introduce two significant efforts on Astra-sim.

First, we extend ASTRA-sim by integrating detailed modeling of memory usage, computation, and communication overheads for key LLM operators. Additionally, we incorporate a DRAM chiplet capacity model to capture the on-die memory constraints. These enhancements enable memory-aware scheduling simulations, allowing Astra-sim to more accurately reflect the resource consumption on WSCs.

Secondly, to enable rapid simulations, we pre-build a mapping lookup table offline by profiling optimal intra-die mapping strategies for representative configurations, including different models, sequence lengths, and batch sizes. For each case, we log key metrics such as latency, DRAM access volume, and communication overhead. During WATOS’s online execution, the table is accessed in a read-only manner with negligible overhead. The feasibility of this approach is supported by numerous works \cite{jia2019optimizing}, showing the error of precomputed results remains within an acceptable range.

\begin{table}[t]
\renewcommand{\arraystretch}{1.05}
\centering
\caption{Parameters of four Representative Hardware Configurations}\vspace{-1mm}
\begin{tabular}{l|c|c|c|c}
\specialrule{0.1em}{0.5pt}{0.98pt}
\textbf{Config. Type} & \textbf{Config 1} & \textbf{Config 2} & \textbf{Config 3} & \textbf{Config 4} \\
\specialrule{0.12em}{0.5pt}{0.9pt}
\#Die num & 64 & 56 & 56 & 48 \\
\#Die in X,Y-dim & (8,\,8) & (7,\,8) & (7,\,8) & (6,\,8) \\
Comp Power & 512 & 708 & 708 & 708 \\
DRAM Bandwidth & 1TB/s & 1.5TB/s & 2TB/s & 2.5TB/s \\
DRAM per die & 48GB & 64GB & 70GB & 96GB \\
D2D Bandwidth & 4.5TB/s & 4.5TB/s & 4TB/s & 3.5TB/s \\
\specialrule{0.12em}{0.5pt}{0.9pt}
\end{tabular}
\label{tab:wafer_config_params}
\end{table}

\section{Evaluation}\label{sec:Evaluation}
\subsection{Experiment Setup}\label{subsec:set_up}

\emph{\textbf{Basic Hardware Configurations}}. To support wafer-scale architecture exploration, we configure the compute die parameters based on the setup illustrated in Fig. \ref{fig:wafer}. The compute die have two configurations: (1) 21.92mm × 22.81mm with a 16$\times$16 array of Dojo-style cores, and (2) 25.5mm × 25.2mm with an 18×18 core array. They are manufactured using TSMC’s 7 nm process technology and operates at a frequency of 2 GHz \cite{talpes2022dojo}. The edge of the compute die provides a total interconnect bandwidth of 12 TB/s across four directions. Each core delivers a computational power of 2.04 FP16 TFLOPS and features 1.25 MB of SRAM.

\emph{\textbf{Architectural DSE}}. In this study, we perform an architectural DSE of WSCs based on the compute chips. Each chip can integrate a varying number of DRAM chiplets, enabling different trade-offs in compute capability, memory capacity, and bandwidth. This exploration aims to demonstrate the advantages of WATOS and to provide deeper insights into the resource trade-offs on WSCs. Table \ref{tab:wafer_config_params} summarizes four representative wafer-scale configurations generated by the proposed enumerator, which serve as the basis for our DSE.

%\emph{DSE Time}. 

%Gemini: The DSE time increases with the increase in target computing power. For example, the DSE for 72 TOPs and 512 TOPs use 80 threads and 100 threads, respectively, and run for 2280s and 23907s on an Intel Xeon Platinum 8260 server.

% \revlink{revE5}{E-5} \sethlcolor{RE}\hl{\emph{Global Optimization Settings}. The global optimization time increases with both the number of exploration steps and the target computing power. For exploration steps, the termination condition is set as $Performance^{(k+\Delta T)} - Performance^{(k)} \le \epsilon$. As depicted in Fig. XX, under different workloads, the average number of steps required to meet this convergence criterion is XX, which is used as the exploration steps. Regarding computing power, the global optimization run for 23,907 s on an Intel Xeon Platinum 8260 server with 64 threads.}

\emph{\textbf{DSE Configurations}}. We evaluate WATOS on Intel Xeon Gold 6448H with 32 cores per socket. The runtime of the global optimizer is 0.274s for 100 exploration steps. Beyond runtime, we observe that the selection strategy inherently introduces a trade-off between exploration steps and final performance: 1) The elitist strategy selects only the fittest individuals, leading to fast convergence but often suboptimal performance. 2) Binary tournament selection preserves greater diversity, enabling better performance at the cost of slower convergence. We introduce $\omega$ as an interface to control the proportion of elitism. Fig. \ref{fig:Scaled_Perf} (b) illustrates the impact of varying $\omega$ on both the convergence speed and the final performance.

% The proportion of elitism is controlled by $\omega$. A larger $\omega$ accelerates convergence but increases the risk of suboptimal local optima, whereas a smaller $\omega$ slows convergence and can lead to better performance. This parameter is provided as a user-configurable interface.

%The experiment is executed on a 14-core Intel Core i7-12700H processor utilizing four-thread parallel processing, which completed 100 iterations in 0.2741 seconds. Although consistently selecting individuals with higher fitness in each iteration expedites convergence, the consequent reduction in population diversity increases the likelihood of converging to a local optimum. Therefore, the binary tournament selection strategy is implemented. This method involves the random selection of two individuals, with the individual exhibiting superior fitness advancing, thereby affording suboptimal individuals a probabilistic opportunity for survival. Nevertheless, relying solely on binary tournament selection presents a risk of failing to propagate optimal individuals to the offspring population. To counteract this limitation, an elitist strategy was incorporated, which explicitly preserves a predetermined fraction of the most fit individuals for inclusion in the offspring population.

\emph{\textbf{Workloads}}. We choose a series of models from 30B to 671B for our experiments, including dense models Llama2-30B \cite{touvron2023llama2}, Llama3-70B \cite{llama3modelcard}, GPT-175B \cite{brown2020language} and MoE models Gshard-137B \cite{lepikhin2020gshard} and Deepseek-v3 \cite{deepseekai2025deepseekv3technicalreport}. We adopt mixed precision training, with FP16 for weights and activations,
and FP32 for Adam optimizers.

% \revlink{revD5}{D-5} \sethlcolor{RD}\hl{Notably, for the case that dimensions cannot be evenly divisible by the number of partitions. WATOS rounds up the tensor shape to a multiple of the partition count, padding with operator-specific values to ensure correctness.}

\emph{\textbf{Metrics}}. To evaluate system performance, we adopt iteration time and throughput as metrics. Iteration time measures the latency of one forward and backward pass, while throughput denotes floating-point operations per second. All results are normalized to the lowest-performing configuration (value = 1). As recomputation may increase throughput without improving performance due to additional computation, we break down throughput results for all configurations using recomputation.

\subsection{Architectural Design Space Exploration}\label{subsec:Archi_DSE}

\begin{figure}[t]
\includegraphics[width=\linewidth]{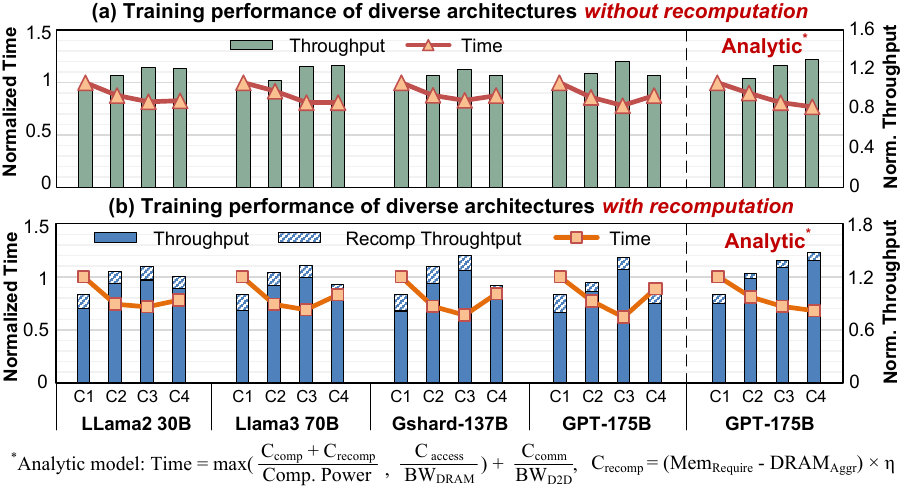}\vspace{-3mm}
\caption{Performance comparison for configs 1-4 on Llama2 30B, Llama3-70B, Gshard-137B, GPT-175B  (a) w/o recomputation (b) w/ recomputation. Analytic model for DSE optimum only; idealized results not comparable.}
\label{fig:case}
\end{figure}

As recomputation is critical in improving memory efficiency during training, a viable WSC must perform robustly under both recomputation and non-recomputation scenarios. Fig.\ref{fig:case} shows results of four feasible WSC architectures across representative LLM models, covering variations in sequence length and batch size for generalizable evaluation. Config 3 consistently achieves the best performance, demonstrating adaptability to diverse models and workloads, and emerging as the universal optimal WSC. These results suggest that a WSC with moderate per-die DRAM capacity and high compute density can effectively balance compute, memory, and communication demand during LLM training. In contrast, architectures with excessively large or small per-die DRAM suffer from imbalanced utilization and degraded performance.

The comparison between Config 2 and Config 4 shows that Config 4 performs better without recomputation, whereas Config 2 excels when recomputation is enabled. This shift stems from recomputation increasing compute demand while reducing DRAM requirements. As a result, Config 2's higher compute density becomes beneficial under recomputation, but leads to performance degradation in non-recomputation scenarios. \emph{This observation underscores that compute density, along with DRAM bandwidth and capacity, are key determinants of performance in LLM training workloads. }

We further apply a first-order analytic model to GPT-175B, as depicted in Fig. \ref{fig:case}, where C$_{\rm comp}$, C$_{\rm recomp}$, C$_{\rm access}$ and C$_{\rm comm}$ denotes the computation load, recomputation load, DRAM access and communication load, respectively. $\eta$ represents the computation load associated with each byte of data. As can be seen, this model fails to capture the aforementioned insights and consistently favors configs with the largest DRAM capacity. This is essentially because the knapsack-like problem solved by WATOS is NP-hard and typically requires either complex multi-dimensional dynamic programming \mbox{\cite{martello1990knapsack, lodi1999approximation}} or heuristic algorithms \mbox{\cite{hiremath2008new, chu1998genetic, lin1993applying}}. This again confirms the necessity of using WATOS for exploring wafer-scale architectures under realistic training conditions.

% \mbox{\cite{chekuri2005polynomial, hartmanis1982computers}}

\subsection{Overall Performance}\label{subsec:Overall_Perf}
We compare the optimal WSC configuration (Config 3) with the Megatron-LM \cite{shoeybi2019megatron}, a SOTA GPU-based LLM training system, referred to as MG-GPU. To ensure a fair comparison, the Megatron setup consists of 8×Blackwell Ultra 288GB GPUs, connected via 1.8 TB/s intra-node bandwidth. In terms of compute resources, MG-GPU offers a total of 40,000 TFLOPS, slightly exceeding the 39,648 TFLOPS provided by the WSC. To ensure fairness, MG-GPU’s DRAM is scaled from 2304 GB to 3920 GB to match the WSC, while both systems maintain equal DRAM bandwidth of 2 TB/s. Also, we directly apply Megatron’s scheduling strategy and Cerebras weight streaming strategy to the WSC, denoted as MG-wafer and Cerebras, respectively. For MG-wafer, we determine TP and PP sizes using Megatron, enumerate all feasible physical shapes on wsc (e.g., 1×4, 2×2, 4×1 for TP=4), place them in the naive serpentine arrangement (Fig. \ref{fig:placement}(a)) and evaluate the performance. We pick the best config as results for MG-wafer. 

\begin{figure}[t]
\centering 
\includegraphics[width=\linewidth]{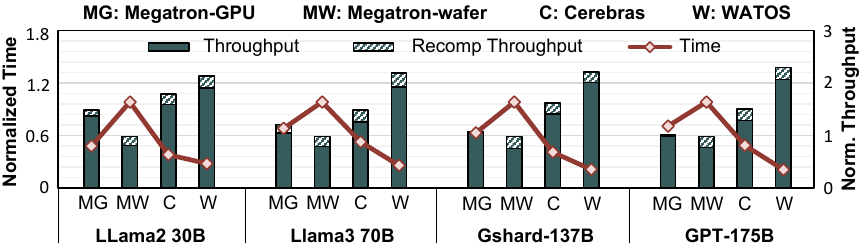}\vspace{-2mm}
\caption{Overall performance comparisons among Megatron-GPU, Megatron-wafer, Cerebras and WATOS.}
\label{fig:overall_performance}\vspace{-2mm}
\end{figure}

Fig. \ref{fig:overall_performance} compares the normalized throughput and training time across various LLM models. Compared to MG-GPU, WATOS achieves a $1.92\times$ higher throughput and $1.75\times$ shorter iteration time. Against MG-wafer and Cerebras, WATOS achieves up to 2.74$\times$ and 1.53$\times$ higher throughput, along with 2.45$\times$ and 1.31$\times$ shorter iteration time, respectively. Despite fewer compute resources, WATOS outperforms due to efficient operator execution and training-aware scheduling tailored to wafer-scale architectures. The performance gap between MG-wafer and MG-GPU stems from a $23\%$ reduction in memory resources on MG-wafer and Megatron’s scheduling policy is optimized for GPU clusters, lacking the hardware-aware refinements required for wafer-scale systems. Notably, the proportion of recomputation in WATOS is only $30\%$ to $60\%$ of that in MG-Wafer, exhibiting the effectiveness of the proposed memory-balancing scheduling to reduce pipeline bubbles caused by recomputation and promote training efficiency. Compared to Cerebras, the iteration time reduction mainly stems from lower communication overhead and becomes more pronounced under small batch sizes and short sequences, up to a $187.4\%$ improvement, because the communication cost of weight streaming scales proportionally with model parallelism degree. These results underscore the effectiveness of WATOS on WSCs for efficient LLM training.

% \begin{figure}[t]
% \centering\vspace{-2mm}
% \includegraphics[width=\linewidth]{figures/Ablation.pdf}\vspace{-4mm}
% \caption{Ablation study for different optimizations of WATOS.}
% \label{fig:Ablation_study}\vspace{-2mm}
% \end{figure}

\begin{figure}[t]
\centering
\includegraphics[width=\linewidth]{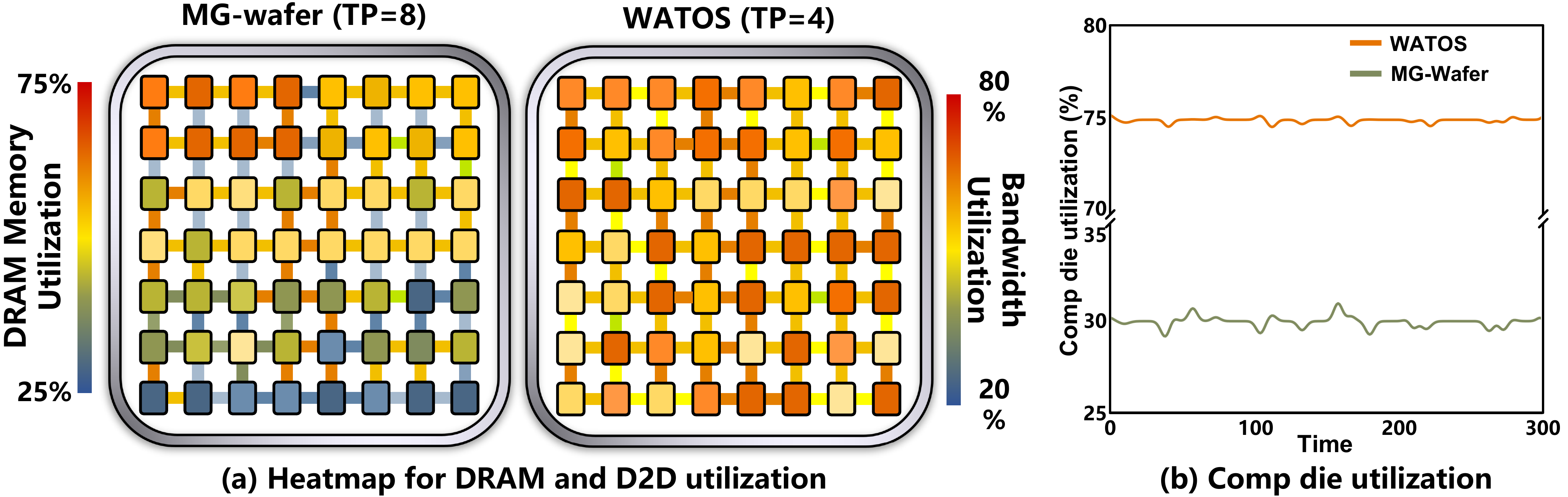}\vspace{-2mm}
\caption{DRAM memory, D2D and computation utilization comparisons between MG-wafer and WATOS for GPT 175B.}
\label{fig:utilization}
\end{figure}

\begin{figure}[t]
\centering
\includegraphics[width=\linewidth]{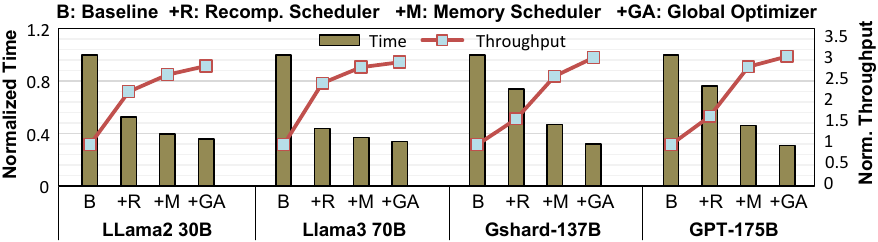}\vspace{-2mm}
\caption{Ablation performance of the WATOS.}
\label{fig:Ablation_study}\vspace{-2mm}
\end{figure}

\begin{figure}[t]
\centering
\includegraphics[width=\linewidth]{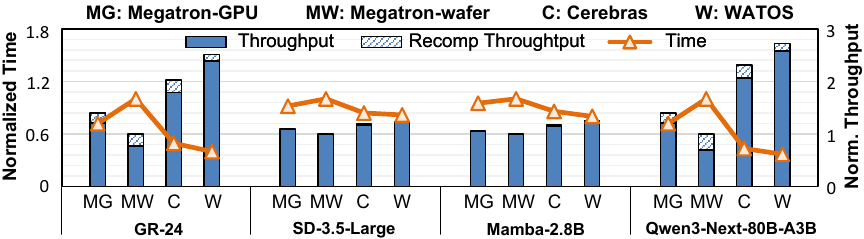}\vspace{-3mm}
\caption{WATOS's performance on different emerging models.}
\label{fig:modelgenerality}
\end{figure}

\textbf{Resource Utilization.} Fig. \ref{fig:utilization} presents the heatmap under GPT-175B. As shown, WATOS with TP=4 achieves higher DRAM utilization and bandwidth efficiency than MG-wafer with TP=8. This is because MG-wafer suffers from inefficient resource partitioning and placement, leading to longer communication paths and frequent congestion that stalls computation. As a result, the compute die utilization in MG-wafer is only about $40\%$ of that in WATOS. This highlights an insight: \emph{With efficient resource allocation and memory scheduling strategies, small tensor parallelism can exhibit better performance on WSCs. }

\begin{figure}[t]
\centering 
\includegraphics[width=\linewidth]{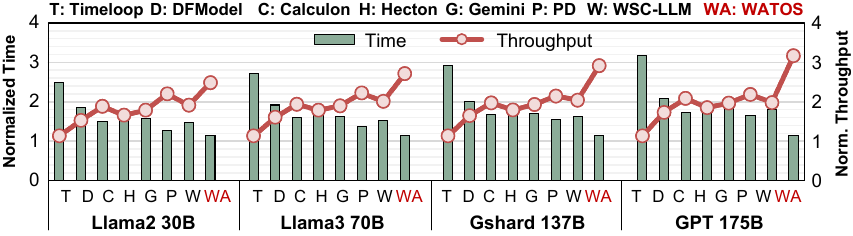}\vspace{-2mm}
\caption{Performance comparisons among WATOS and DSE methods.}
\label{fig:dse}
\end{figure}

\begin{figure}[t]
\centering
\includegraphics[width=\linewidth]{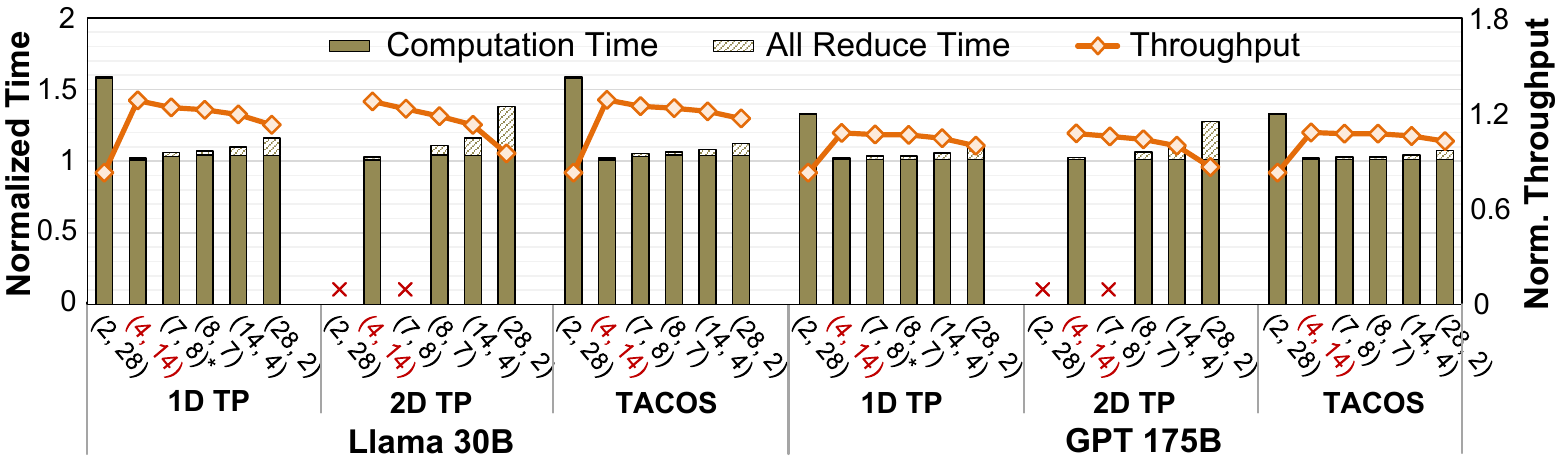}\vspace{-2mm}
\caption{Performance of different TP strategies and (TP, PP) configs.}
\label{fig:HiDimTP}\vspace{-2mm}
\end{figure}

\subsection{Ablation Experiments}\label{subsec:Ablation_Experiemnt}
To assess the impact of WATOS optimizations, we conduct an ablation study on the 56-die wafer (Config 3). The baseline uses naive parallelism with fixed TP=8 and PP=7, without WATOS's enhancements. Optimizations are then incrementally applied: \textbf{+R} enables GCMR recomputation scheduler, \textbf{+M} adds a location-aware memory scheduler, \textbf{+GA} incorporates the GA-based global optimizer.

As depicted in Fig.\ref{fig:Ablation_study}, we make the following observations: (1) As the model size increases, the performance benefit of the centralized scheduler diminishes. (2) In contrast, the performance gains from the memory-aware scheduler and recomputation grow with larger models. This is because smaller models typically require fewer pipeline stages, resulting in relatively mild memory imbalance issues and thus limited opportunities for memory scheduling optimization. In contrast, as model size increases, the pipeline depth grows, making memory scheduling critical for maintaining computational efficiency. Moreover, larger models are more susceptible to memory overflow, which amplifies the benefits of recomputation-based scheduling. These results highlight that the coordinated use of memory-aware, and recomputation schedulers in WATOS enables efficient training across diverse LLM scales.

\section{Discussion}\label{sec;discussion}
\subsection{DSE Methods Analysis} 
Fig. \mbox{\ref{fig:dse}} compares the performance of seven representative DSE studies for training LLMs under the WSC setting. We reproduce these works and capture their performance on WSC, including the DNN accelerator DSE framework Timeloop \mbox{\cite{parashar2019timeloop}}, the training-aware optimization DFModel \mbox{\cite{ko2024dfmodel}}, Calculon
\mbox{\cite{isaev2023calculon}}, two chiplet-based DSE frameworks: Hecton \mbox{\cite{huang2024hecaton}}, Gemini \mbox{\cite{cai2024gemini}}, and two wafer-scale DSE schemes: PD \mbox{\cite{yang2025pd}}, WSC-LLM \mbox{\cite{xu2025wsc}}. Five key insights are made: \textbf{1)} Timeloop performs worst as it targets only core-level workload partitioning, neglecting DRAM-aware and inter-die optimizations. \textbf{2)} DFModel and Calculon partially improve performance by considering multi-dimensional parallel spaces. Calculon’s additional use of training memory-saving techniques provides more performance gains. \textbf{3)} Hecaton and Gemini fully account for communication in a 2D mesh, but as chiplet-oriented designs, they focus on reducing DRAM accesses rather than DRAM capacity, which leads to suboptimal performance. Moreover, Hecaton’s 2D TP relies on bypass links, resulting in additional communication costs on a 2D mesh topology. \textbf{4)} PD explores optimal topologies and logical designs in topological spaces but does not alleviate DRAM scarcity on WSCs, and its interconnect-focused design can worsen memory constraints. \textbf{5)} WSC-LLM explores the optimal LLM inference architecture under area constraints; however, its design space is suboptimal due to the lack of recomputation-aware optimization. \textbf{Overall}, WATOS achieves superior performance by jointly considering WSC topology constraints, area trade-offs, and memory balance.

\subsection{Expansion of Parallelism Seach Space} 

Fig. \ref{fig:HiDimTP} demonstrates WATOS’s performance in the expanded parallelism search space, including 2D TP proposed by GSPMD \mbox{\cite{xu2021gspmd}}, RingBiOdd \cite{laskar2024enhancing} and TACOS \cite{won2024tacos}. The WSC configuration remains consistent with \texttt{Config3}. Notably, RingBiOdd is employed to implement 7-instance TP that cannot be handled by the naive AllReduce \mbox{\cite{chen2024rina}}.

Fig. \mbox{\ref{fig:HiDimTP}} illustrates two key insights: \textbf{1)} The expanded parallel search space does not alter the optimal design point. Although TACOS and RingBiOdd support odd TP instances and TACOS outperforms at larger TP sizes, they still cannot overcome the increasing communication overhead as TP size grows. \textbf{2)} 2D TP yields the worst performance on the 2D mesh topology. While 2D TP can decompose large TPs, it is not tailored for LLMs, leading to higher communication volume and significant tail latency on a 2D mesh. Collectively, 2D TP, RingBiOdd, and TACOS expand WATOS’s parallel search space while leaving the optimal design point unchanged.

\begin{figure}[t]
\centering
\includegraphics[width=\linewidth]{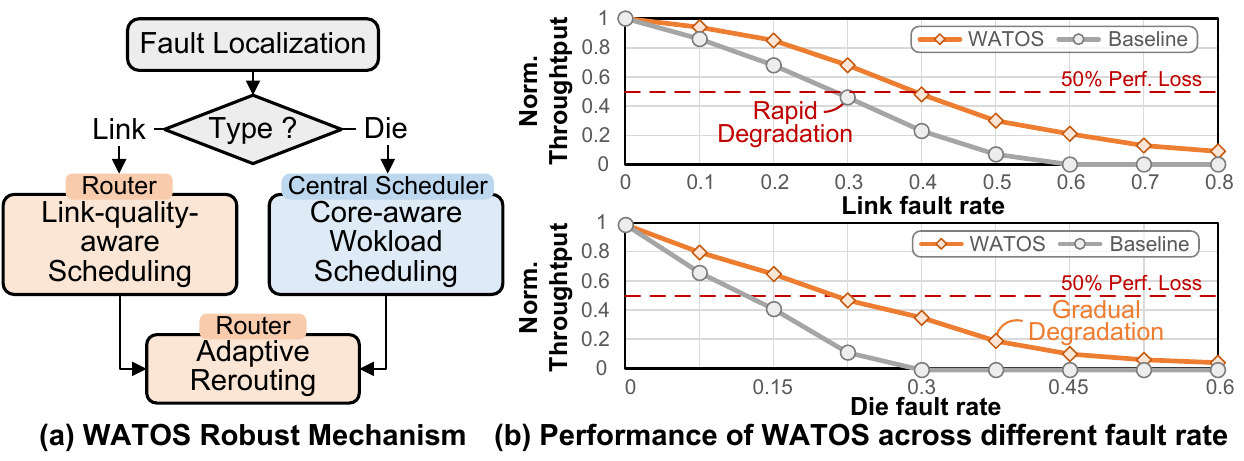}\vspace{-2mm}
\caption{The robustness and reliability design of WATOS.}
\label{fig:robust}
\end{figure}

\begin{figure}[t]
\centering
\includegraphics[width=\linewidth]{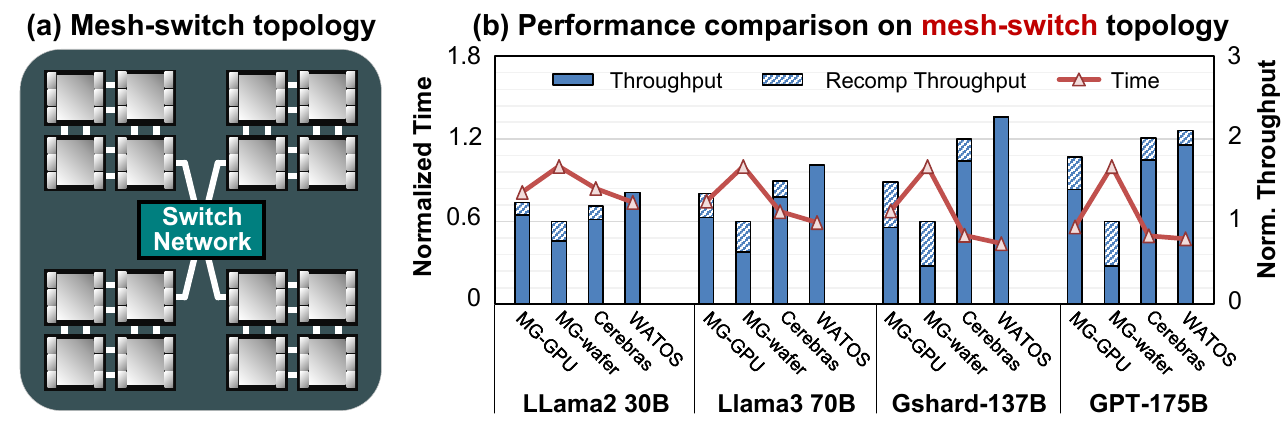}\vspace{-2mm}
\caption{WATOS performance on Mesh-switch topology.}
\label{fig:topology}
\end{figure}

\subsection{Architectural Generality} 
Fig. \mbox{\ref{fig:modelgenerality}} demonstrates remarkable generality of WATOS across emerging models, including Generative Recommender \mbox{\cite{zhai2024actions}} for recommendation, Stable Diffusion \mbox{\cite{esser2024scaling}} for image and video generation, as well as new LLM architectures such as Mamba \mbox{\cite{gu2024mamba, dao2024transformers}} and Qwen3-Next \mbox{\cite{qwen3next}}, which are based on state space models and linear gated attention, respectively. The hardware configuration remains consistent with \texttt{Config 3}.

The strong generality of WATOS stems from three key factors. 1) WATOS offers configurable architectural parameters to enable adaptation to diverse wafer-scale hardware setups and performs end-to-end scheduling to optimize large-scale model training. 2) WATOS is operator-centric rather than model-specific, allowing it to accommodate new model workloads by extending DNN predictor, TP engine and compute die configurations without hardware modifications whenever new operators emerge. 3) WATOS targets structural characteristics common to large-scale neural network training, addressing memory imbalance across pipeline stages and employing adaptive recomputation strategies rather than relying on any specific LLM or hardware architecture. Overall, WATOS demonstrates strong generality in large-scale model training rather than being a niche framework tailored for a niche architecture.

\subsection{Robustness and reliability} 
As shown in Fig. \mbox{\ref{fig:robust}} (a), WATOS introduces a 3-stage system-level robustness design: \textbf{1)} fault localization and classification, \textbf{2)} link-quality- and core-aware workload scheduling, and \textbf{3)} adaptive rerouting. \textbf{For link faults}, the router continuously monitors the quality of its connected links, such as packet loss rate and peak D2D bandwidth. Upon detecting degradation, it proactively reduces the communication load on the affected links and reroutes the excess traffic to the nearest alternative path using well-established routing algorithms \mbox{\cite{niranjan2009portland, pignolet2017load}}. In the rare case of a complete link failure, no traffic is allocated to the broken link. \textbf{For die faults}, prolonged operation may lead to core degradation, resulting in reduced die performance. The central scheduler dynamically adjusts workload distribution based on the degree of degradation, assigning less workload to the faulty die and redistributing tasks to healthy ones to maintain computation balance. Die failures, rare but critical, trigger the WATOS to mark the faulty die and its links as unavailable, excluding them from further workload allocation.

To evaluate fault-tolerance performance, we conduct experiments on WSC Config 3 across diverse LLM workloads, using the non-robust version of WATOS as the baseline and manually injecting failures. As depicted in Fig. \mbox{\ref{fig:robust}}, under a 20\% link fault rate and a 20\% die fault rate, the fault-tolerant WATOS achieves 18\% and 35\% higher throughput, respectively, benefiting from the system-level robustness optimization.

\subsection{Topology Compatibility} 
Fig. \mbox{\ref{fig:topology}} evaluates the performance of WATOS on mesh-switch topology proposed by \mbox{\cite{yang2025pd}}. As illustrated in Fig. \mbox{\ref{fig:topology}} (a), we reconfigure \texttt{Config3} to 48 dies arranged in a 12×2×2 mesh with a switch network offering 1.6 TB/s of bandwidth. The GPU baseline is scaled to ensure equivalent compute capability. Fig. \mbox{\ref{fig:topology}} (b) illustrates that WATOS maintains superior performance compared to Megatron and Cerebras on the Mesh-switch topology, because: \textbf{1)} WATOS restricts TP within the mesh to fully leverage its high bandwidth and routes the lightweight inter-stage communication via the switch; \textbf{2)} the flexible switch enhances communication within mem\_pair, alleviating potential congestion. By contrast, Megatron’s lack of mesh awareness leads to oversized TP and switch-limited All-Reduce, while Cerebras’ weight streaming similarly underutilizes mesh bandwidth and remains switch-bound. In such case, WATOS sustains high performance on mesh-switch topologies, demonstrating strong compatibility across diverse interconnect topologies.

\textbf{3D Stacking variants.} The 3D stacking \mbox{\cite{li2025h2, li20253d, zhao20253d}} which vertically places DRAM on top of compute dies fundamentally shifts the memory access pattern and architectural trade-offs on WSC: \textbf{1)} DRAM access exhibits a local-remote gap. Local DRAM accesses traverse only the hybrid bonding interface, whereas remote accesses to DRAM on other compute dies must go through the NoC. \textbf{2)} 3D stacking decouples area competition between DRAM and compute dies, which allows fully allocating IOs to D2D interconnects, enabling larger DRAM capacity, higher compute density and increased D2D bandwidth across the wafer. Nonetheless, 3D stacking does not fundamentally alter the intrinsic characteristics of LLM training on WSCs: \textbf{1)} Communication relies on a high-bandwidth 2D mesh, which achieves high utilization for small TP size. \textbf{2)} Memory imbalance across pipeline stages necessitates recomputation. \textbf{3)} Unbalanced recomputation introduces bubbles, degrading performance. Thus, WATOS still remains competitive under 3D stacking, with a more relaxed trade-off.

\begin{figure}[t]
\centering
\includegraphics[width=\linewidth]{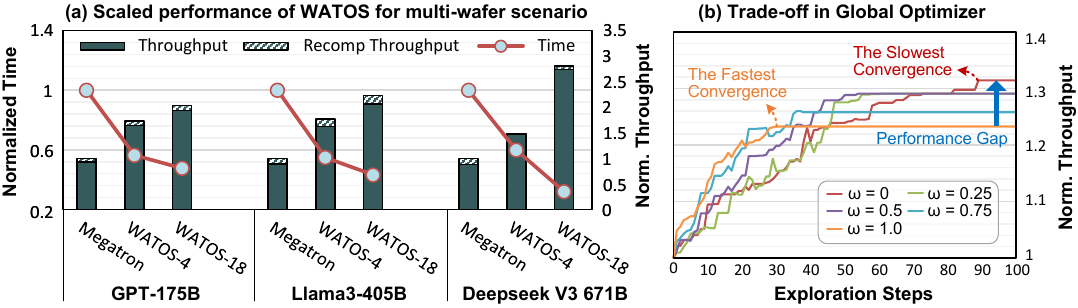}\vspace{-2mm}
\caption{(a) WATOS performance on multi-wafer scenarios. (b) Time-performance trade-off in the global optimizer.}
\label{fig:Scaled_Perf}
\end{figure}

\subsection{Scalability for Multi-WSC Scenarios}
We construct a WSC node composed of four Config-3 wafers and compare its performance against Megatron \cite{shoeybi2019megatron} to evaluate the scalability of WATOS, which comprises four nodes, each containing eight NVIDIA Blackwell ultra GPUs. For wafer-to-wafer (W2W) interconnect bandwidth, we consider two configurations: 1) WATOS-18, which employs a SOTA W2W interconnect \cite{talpes2022dojo} delivering 1.8 TB/s bandwidth; and 2) WATOS-4, which adopts a 400 GB/s W2W bandwidth configuration, equivalent to the inter-node bandwidth of Megatron. Further, we evaluate the capability of WATOS in training ultra-large models, including LLaMA3-405B \cite{llama3_405b} and Deepseek V3-671B \cite{deepseekai2025deepseekv3technicalreport}.

As shown in Fig. \ref{fig:Scaled_Perf} (a), two key insights emerge: 1) WATOS consistently outperforms Megatron in both high- and low-bandwidth W2W configurations; 2) WATOS achieves greater performance gains when scaling to ultra-large LLMs, demonstrating superior scalability. These findings arise from two main factors. For insight (1), WATOS outperforms for its ability to fully utilize the compute and memory resources within a single wafer, thereby reducing the reliance on W2W communication.  For Insight (2), the scalability advantage stems from the high memory demand of large models. For example, Llama3-405B requires around 5670 GB of memory to store model weights, optimizer states, and gradients. As a result, Megatron must distribute the model across at least three 8-GPU servers. In contrast, two WSCs are sufficient, requiring at most a hop cross-wafer communication. Therefore, even WATOS-4 can achieve significant performance gains.

\begin{figure}[t]
\centering
\includegraphics[width=\linewidth]{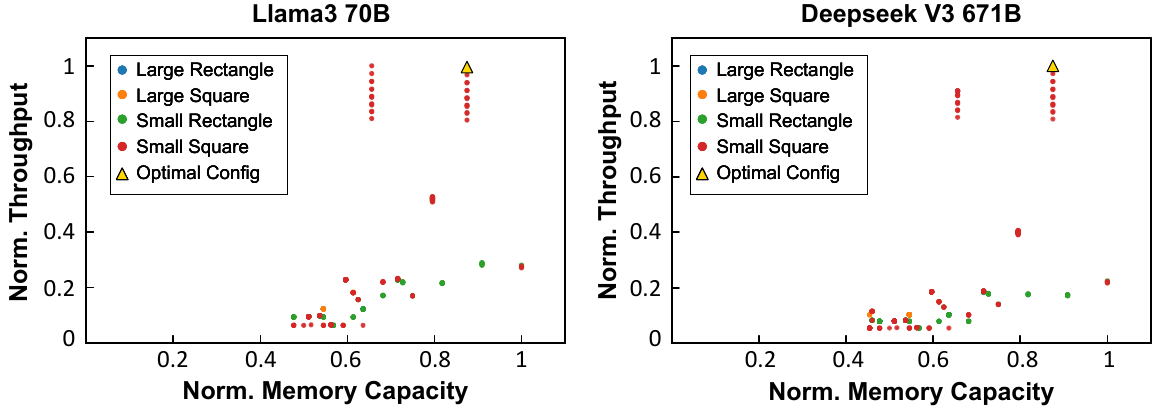}\vspace{-2mm}
\caption{Hardware DSE. Points of the same color denote configurations with the same die size category (Small or Large) and shape (Square or Rectangle), while different markers indicate different size categories and shape settings.}
\label{fig:DSE}
\end{figure}

\textbf{Hardware DSE.} With the proposed hardware template, we explore the design space at die granularity, considering dies from 200\,mm$^2$ to 600\,mm$^2$, categorized as Small ($<$\,400\,mm$^2$) or Large, and as Square (aspect ratio$<$\,1.2) or Rectangle. The DSE objective is the product of memory capacity and throughput. As depicted in Fig.~\ref{fig:DSE}, under diverse workloads, Small Square configurations deliver superior performance by providing longer edges for D2D links and better area utilization, whereas Rectangle layouts underutilize area along single dimension, reducing both throughput and memory capacity. However, the high power density of Small Square configurations imposes strict cooling and power delivery constraints, leading current WSCs to favor Large Square design.

\section{Related Works}
\textbf{LLM training framework.}
Numerous frameworks \cite{shoeybi2019megatron, narayanan2021efficient, korthikanti2023reducing, rasley2020deepspeed, rajbhandari2020zero, zheng2022alpa, li2023colossal, tang2023torchsparse++, gurevin2024prunegnn} have been developed to improve LLM training efficiency on GPUs. For example, Alpa \cite{zheng2022alpa} decomposes the parallelism space into inter- and intra-operator dimensions to generate optimized execution plans, while Colossal-AI \cite{li2023colossal} supports modular parallelism across data, tensor, and pipeline levels. However, these works are tailored to fully-connected GPUs and are not optimized for WSCs with 2D mesh interconnects. Thus, they overlook opportunities to exploit on-wafer bandwidth for fine-grained resource allocation.

% \textbf{Architectural DSE for Die-Level.} Numerous studies \cite{zhu2024theseus, feng2024switch, pal2021scale, chenwaferscale, rashidi2024fred, pal2021designing, hu2024wafer, yu2025cramming, yang2025pd, xu2025wsc, parashar2019timeloop, wu2022sparseloop, horeni2022ruby, mei2023defines, hong2023dosa, yang2020interstellar} have been devoted to architectural exploration. For example, Timeloop \mbox{\cite{parashar2019timeloop}} develop a workload mapping-exploration framework to systematically optimize DNN accelerator designs. Sparseloop \mbox{\cite{wu2022sparseloop}} extends sparse workloads, while Ruby \mbox{\cite{horeni2022ruby}} further handles imperfectly-factorized mappings. Regarding DSE methods, many works \mbox{\cite{giordano2024tinyforge, bi2023heron, wang2025using, ren2024enabling, dat2025hsevo, karystinos2024harpocrates, sun2020automatically}} employ genetic algorithms. Harpocrates \mbox{\cite{karystinos2024harpocrates}} uses program mutation operators to automatically generate efficient test programs, while \mbox{\cite{sun2020automatically}} proposes a variable-length encoding scheme with tailored operators to optimize CNN architectures. These operators are task-specific and difficult to generalize across training tasks on WSC. 

\textbf{Architectural DSE for Die-Level.} Numerous studies \cite{parashar2019timeloop, wu2022sparseloop, horeni2022ruby, mei2023defines, hong2023dosa, yang2020interstellar} have been devoted to architectural exploration. For example, Timeloop \mbox{\cite{parashar2019timeloop}} develop a workload mapping-exploration framework to systematically optimize DNN accelerator designs. Sparseloop \mbox{\cite{wu2022sparseloop}} extends sparse workloads, while Ruby \mbox{\cite{horeni2022ruby}} further handles imperfectly-factorized mappings. Regarding DSE methods, many works \mbox{\cite{giordano2024tinyforge, bi2023heron, wang2025using, ren2024enabling, dat2025hsevo, karystinos2024harpocrates, sun2020automatically}} employ genetic algorithms. Harpocrates \mbox{\cite{karystinos2024harpocrates}} uses program mutation operators to automatically generate efficient test programs, while \mbox{\cite{sun2020automatically}} proposes a variable-length encoding scheme with tailored operators to optimize CNN architectures. These operators are task-specific and difficult to generalize across training tasks on WSC. 

\textbf{For communication DSE}. Several works \mbox{\cite{rashidi2022themis,farrokhbakht2022stay,tong2024feather,rashidi2024fred,won2024tacos,laskar2024enhancing,ko2024dfmodel,isaev2023calculon}} explore the design space of parallelism and network topology. FRED \mbox{\cite{rashidi2024fred}} investigates the optimal switch topology and corresponding parallelization strategies at the wafer scale. TACOS \mbox{\cite{won2024tacos}} searches for optimal collective communication algorithms for a given topology by leveraging a time-expanded network representation and a link-chunk matching algorithm. Nonetheless, these approaches lack consideration of DRAM capacity during LLM training. 

\textbf{For chiplets}. Current studies investigate architecture \cite{zhao2023cooco,orenes2024muchisim,cai2024gemini,zhang2024llmcompass,kadiyala2024leveraging}, dataflow \cite{gao2019tangram,orenes2023dcra} and SRAM optimizations \cite{shao2019simba,garg2024pipeorgan}. However, these works focus on DNN inference in the chiplet scale, whereas our focus is on optimizing LLM training efficiency in WSCs. \textbf{For WSCs}, existing works \cite{he2025waferllm,zhu2024theseus, feng2024switch, pal2021scale, chenwaferscale, rashidi2024fred, pal2021designing, hu2024wafer, yu2025cramming, yang2025pd, xu2025wsc, wang2024tmac,wei2025spatial,yang2025segmentation,li2024research,wan2024architectural} have explored the design space of wafer-scale architectures, as well as floorplan design \cite{jiang2021cu,ozdemir2022kernel,luo2023ms,liu2022partition}. However, existing works predominantly focus on inference, leaving training strategy co-design largely unexplored. While TMAC \cite{wang2024tmac} pioneers the exploration of resource trade-offs under constrained wafer-scale area for recomputation, it lacks a global optimization strategy, which makes it difficult to identify the optimal design point. In summary, to the best of our knowledge, no prior work has systematically investigated optimal parallelism strategies and resource allocation for LLM training on wafer-scale systems, making WATOS the first to address this problem.

% WSC-LLM \cite{xu2025wsc} explores LLM inference strategies under WSCs resource constraints.

\section{Conclusion}
This paper presents WATOS, an efficient framework for co-exploring LLM training and architecture on WSCs. Leveraging WSCs' high D2D bandwidth and unique topology, WATOS optimizes parallelism, recomputation scheduling, placement, and memory allocation to improve LLM training performance. Experiments show that WATOS achieves $2.74\times$ throughput gains over Megatron and $1.53\times$ than Cerebras. 

\section*{Acknowledgments}
This work was supported in part by the National Science and Technology Major Project under Grant 2022ZD0115200; the NSFC under Grant 62125403, and Grant 92164301; Beijing S\&T Project Z221100007722023; in part by the project funding for the 2022 Special Project on Industrial Foundation Reconstruction and High Quality Development of Manufacturing Industry CEIEC-2022-ZM020245; in part by the Beijing National Research Center for Information Science and Technology; and in part by the Beijing Advanced Innovation Center for Integrated Circuits.

%%%%%%% -- PAPER CONTENT ENDS -- %%%%%%%%

%%%%%%%%% -- BIB STYLE AND FILE -- %%%%%%%%
\bibliographystyle{IEEEtranS}
\bibliography{main}
%%%%%%%%%%%%%%%%%%%%%%%%%%%%%%%%%%%%

\end{document}